\documentclass[journal]{IEEEtran}
\usepackage{bm}
\usepackage{setspace}
\usepackage{cite}
\usepackage{amsthm}
\usepackage{epsfig}
\usepackage{amsmath}
\usepackage{amssymb}
\usepackage{latexsym}
\usepackage{enumitem}
\usepackage{mathtools}
\usepackage{amsfonts}
\usepackage{color}
\usepackage{optidef}
\usepackage{lettrine}
\usepackage{graphicx}
\usepackage{multicol,flushend,color}
\usepackage{wasysym}
\usepackage{float}
\usepackage{algorithm}
\usepackage[noend]{algpseudocode}
\usepackage{xcolor}
\usepackage{cuted}
\usepackage{url}
\usepackage[font=footnotesize]{caption}
\usepackage[font=footnotesize]{subcaption}

\newcommand{\E}{\ensuremath{\mathbb E}}
\def \treq {\stackrel{\tiny \Delta}{=}}

\begin{document}

	\title{Deep Joint Source Channel Coding for {Privacy-Aware} End-to-End  
		 Image Transmission}
	\markboth{}{}
	\author{}
	\author{\IEEEauthorblockN
		{	 
			Mehdi Letafati$^{\ast}$, \IEEEmembership{Student Member, IEEE,} 
		Seyyed 
   Amirhossein Ameli Kalkhoran$^{\ast}$, 
			Ecenaz Erdemir,  \IEEEmembership{Student Member, IEEE,} 
			Babak Hossein Khalaj, 
			\IEEEmembership{Senior Member, IEEE,}   
			Hamid Behroozi, \IEEEmembership{Member, IEEE,} 
			\\ and
			Deniz G\"{u}nd\"{u}z,
			\IEEEmembership{Fellow, IEEE}
		}
		\textsuperscript{}
		\thanks{ 
			${}^\ast$ Equal contribution.   

              Some preliminary results of this work were presented, in part, at IEEE Global Communications Conference, Kuala Lumpur, Malaysia, Dec. 2023 \cite{Deep_JSCC_GC}. 
			
			M. Letafati 
was   with the Electrical Engineering Department of 
			Sharif University of Technology, Tehran,  
		1365-11155,
			Iran,  at the time of writing this paper (e-mail:‌ mletafati@ee.sharif.edu).  
            A. Ameli was with the Computer Engineering Department of Sharif University of Technology, Tehran,  
			Iran, at the time of writing this paper (e-mail:‌ ameli@ce.sharif.edu).  
			E. Erdemir 
    was with the Department of Electrical and Electronic Engineering, Imperial College London, London SW7 2AZ, U.K. at the time of writing this paper.
			B. H. Khalaj, 
			and H. Behroozi
			are with the Electrical Engineering Department, 
			Sharif University of Technology, Tehran,  
			Iran (e-mails:   
			khalaj@sharif.edu, behroozi@sharif.edu).  
 D. G\"{u}nd\"{u}z is with the
			Department of Electrical and Electronic Engineering, Imperial College London, London SW7 2AZ, U.K. (e-mail: d.gunduz@imperial.ac.uk).
		}
	}
	
	\IEEEaftertitletext{\vspace{-2\baselineskip}}
	
	\maketitle

	\begin{abstract} 
Deep neural network (DNN)-based joint source and channel coding is proposed for privacy-aware end-to-end  
		image transmission  against
		{multiple 
			eavesdroppers.}
        Both scenarios of {colluding} and {non-colluding} eavesdroppers are considered.  	
   \textcolor{black}{Unlike prior works that assume perfectly known and
independent identically distributed (i.i.d.) source and channel statistics, the
proposed scheme operates under unknown and non-i.i.d. conditions, making it
more applicable to real-world scenarios.} 
		The  goal 
		is to transmit  images
        with minimum distortion, while simultaneously preventing    
		eavesdroppers
		from
		inferring  certain private attributes  of images.  
	Simultaneously generalizing the ideas of \emph{privacy funnel} and wiretap  coding,  a
		{multi-objective} 
		optimization framework is expressed that characterizes the   trade-off between  image reconstruction   quality 
		and information leakage 
        to eavesdroppers, 
	taking into account the  structural similarity index 
			(SSIM)  for improving the perceptual quality of image reconstruction. 
		\textcolor{black}{Extensive experiments on the CIFAR-10 and   
CelebA, along with ablation studies, demonstrate significant performance
improvements in terms of SSIM, adversarial accuracy, and the mutual information leakage compared to benchmarks.}  
Experiments show that the
		proposed 
		scheme   restrains  the  adversarially-trained eavesdroppers from intercepting 
		privatized 
		data for both cases of eavesdropping a common secret,  as well as the case in which eavesdroppers are interested in  different secrets. 
	Furthermore, useful insights on the \emph{privacy-utility trade-off} 
		are also  provided.  
	\end{abstract}

	
	\IEEEpeerreviewmaketitle
	\vspace{-2mm}
	\section{Introduction} 
	\IEEEPARstart{T}{he} 
	sixth generation (6G) of wireless networks is envisioned  to realize  \emph{connected intelligence},  
    supporting  an overabundance of  
    disruptive technologies  such as 
     autonomous driving  and 
    haptic communications for extended  reality (XR)  and metaverse \cite{VR}.   
Recently,  intelligent   \emph{multimedia transmission}
	is receiving much attention  due to its various applications in XR, metaverse, and  surveillance systems \cite{Im-retriev-deniz}.    
	In this regard, if the transmission of  source images is not properly  secured, the performance of such  services are no longer reliable.

Although different radio protocols have been proposed for  the security of core network,  
	the wireless edge of 6G networks seem to face  ever-increasing privacy and security threats, like  man-in-the-middle attacks \cite{twc, WSKG-GC, WSKG-OJVT}, spoofing 
	\cite{WSKG-letter},  and 
	eavesdropping \cite{IoTJ, vtc2022}. 
	Fundamentally, this stems from  the intrinsic 
	open nature of the wireless medium, which makes transmitted signals  susceptible to security and  privacy risks.  
	In this regard, 
	deep neural network (DNN)-based  approaches  
	have shown to be capable of propelling  the performance of wireless systems such that 
	context-aware security can be achieved  
	\cite{ arxive, medical, 6G-PLS}.   	

\subsection{Motivation \& Background}	
		The communication of images from a source node, Alice,  to a legitimate node, Bob, over a noisy wireless channel,  is a joint source channel coding (JSCC) problem. DNN-aided JSCC design, known as DeepJSCC, has received significant attention in  recent years thanks to their superior performance, particularly their lack of reliance on accurate channel state information \cite{IEEE_Proc, DJSCC-Deniz,Deniz-GDN, AE-Deniz, Ecenaz-icassp, deepJSCc_SemCom}.

\textcolor{black}{DeepJSCC frameworks are becoming appealing  for practical deployments in real-world wireless systems \cite{IEEE_Proc}. It is becoming increasingly evident that conventional separate network architectures struggle to meet real-world performance demands, often overlooking signal processing complexities like compression, which can dominate end-to-end latency.  In contrast, DeepJSCC integrates source and channel coding, jointly optimizing them for better overall performance. Moreover, it is seen as a promising solution for reducing bandwidth requirements,  saving radio resources by transmitting only the most relevant information \cite{IEEE_Proc, deepJSCc_SemCom}.}
        
	In JSCC, unlike in separable source and channel coding, the transmitted channel codeword is correlated  with the underlying source signal. While this benefits the legitimate encoder by providing robustness against channel noise, it also creates additional vulnerability in terms of leakage to eavesdroppers. Also note that classical encryption methods are not applicable here as they would destroy the correlation between the source and the channel input.     
    The DNN-based design of JSCC  
    is particularly attractive for the design of secure content delivery schemes, since they can learn the security sensitivity of different parts of the contents, and adopt their transmission strategy  accordingly.   
    Inspired by \cite{AE-Deniz} and \cite{Ecenaz-icassp},
    this paper studies  a generalization of the DeepJSCC approach for privacy-aware  end-to-end image transmission against multiple  eavesdroppers, with both colluding and non-colluding  eavesdropping strategies, and over AWGN as well as  fading channels.  

	
	\vspace{0mm}
	\subsection{Prior Arts} 
	\vspace{0mm}
	In the context of secure end-to-end (E2E)  communications, 
autoencoders were proposed in  \cite{Besser-FFNN, MI-Wiretap-estim, Eduard-AE}  
for communication 
over   AWGN wiretap channel.  Feedforward neural networks composed of linear layers were employed as the encoder-decoder pair,    
	and a  weighted sum of block error rate and approximated information leakage was used as the loss function.   
	The input data to the autoencoder was combined with additional  non-informative random bits 
	to 
	confuse Eve,  while  
    such  redundant bits 
    can negatively affect  the  data   
	rate.   
However, these works only focused on  learning  secure \textit{channel coding} via DNNs rather than  
	taking the source and channel coding jointly into account, i.e.,  undifferentiated  with respect to the \emph{content} of the message being delivered.     

In the context of DNN-based secure wireless image transmission, \cite{DeepJSCC-encryption} proposes cryptography-based approaches (public key cryptography) for secure transmission against one eavesdropper.  It incorporates  DeepJSCC scheme with encryption coding, proposing deep joint source-channel and encryption
coding.  The authors in \cite{DeepJSCC-App}  propose an image ``protection'' scheme with two additional DNN modules (so-called protection and de-protection), parameterized via the U-Net architecture  added to the DeepJSCC pipeline.  
The authors are mainly concerned about 
``application layer'' digital content protection, and they do not take into account the privacy information leakage to  eavesdroppers over wireless links.    
In \cite{SecSemCom},  secure image transmission  with DeepJSCC as the neural model backbone for E2E communication is studied considering only the effect of one eavesdropper. 
This letter assumes the eavesdropper  has access to a similar DNN model as of the legitimate nodes, and  aims to reconstruct images. 
To combat the image reconstruction at the eavesdropper,  
the loss function for training the DeepJSCC pipeline is simply modified by incorporating the mean-squared-error (MSE) between the source images and eavesdropper's reconstruction, which 
  is called ``SecureMSE.''   
Applying SecureMSE for model training  requires 
the statistical characteristics of the eavesdropper's channel to be known by legitimate parties, which  seems not to be feasible in practice, especially in the context of 
``totally passive'' eavesdroppers. 
Recently, a preprint  \cite{SecSemCom-arxiv} 
studies the security of wireless image transmission against multiple eavesdroppers.  The authors approach security from a  different perspective,  i.e.,  signal steganography.   They propose two  additional neural modules to be added to the encoder/decoder, calling it steganography modules.  
Finally, the security of DeepJSCC-based wireless  image transmission against active adversaries and backdoor attacks is studied in \cite{SemCom-AML}, which falls into the category of adversarial machine learning (AML) algorithms. 
    
	\vspace{0mm}
	\subsection{Our Contributions} 
	We propose a DeepJSCC-based solution for secure E2E wireless image transmission  against \emph{multiple eavesdroppers}.  
 By employing DeepJSCC pipeline, idealistic  assumptions of perfectly known and i.i.d. source and channel
distributions are dropped, and the proposed scheme assumes unknown source and
channel statistics.
Generalizing the ideas of privacy funnel and wiretap coding, 
the goal is to transmit images with minimum
distortion, while simultaneously preventing eavesdroppers from
inferring private image attributes, taking into account the perceptual quality metrics for image transmission and reconstruction.    
	We consider both eavesdropping  strategies, i.e.,  the {colluding} setup, in which the eavesdroppers cooperate with each other to extract sensitive/private parts of images, and,  the  {non-colluding} setup, where the eavesdroppers act alone.   
	We also study both cases of eavesdroppers being interested in a common secret, and the case in which every eavesdropper is interested in a different secret.  
	Notably, previous works \cite{AE-Deniz} and \cite{Ecenaz-icassp}  only considered a single eavesdropper with a single-antenna transmitter and multiple parallel channels, respectively. 
	In addition, both  \cite{AE-Deniz} and \cite{Ecenaz-icassp} are limited to static channels, 
	while we assume two well-established  time-varying   wireless channel models. That is,  our proposed  scheme is trained over complex-valued Rayleigh fading channels, and tested over Nakagami-$m$ and AWGN channels in addition to Rayleigh-fading.  
 \textcolor{black}{Our proposed scheme prevents adversarially-trained eavesdroppers from accurately
inferring sensitive attributes of the transmitted images.} 
Moreover,  (different from \cite{Besser-FFNN,Eduard-AE}) no additional 
	redundant bits are required to be added to the source image.    
\textcolor{black}{Unlike the methods in \cite{DeepJSCC-App, 
SecSemCom-arxiv}, our scheme integrates seamlessly into the
standard DeepJSCC pipeline without requiring additional processing modules.}
Instead, we  modify the loss function to come up with a  \emph{privacy-aware} training and inference strategy for the communication entities, taking the information leakage and perceptual quality of images into account.

    Our  scheme relies on a \emph{data-driven} approach, i.e.,   we do not consider  any specific assumption on the underlying  distribution of the data source or the sensitive/private part.  Instead,   we have  access to  datasets to facilitate the process of learning  secure encoder-decoder pairs.
   We provide extensive numerical experiments over  
	 different datasets, i.e.,  CIFAR-10\footnote{\url{https://www.cs.toronto.edu/~kriz/cifar.html}}  and CelebFaces Attributes (CelebA)\footnote{\url{https://mmlab.ie.cuhk.edu.hk/projects/CelebA.html}}. 
	We demonstrate  the performance gain of our approach,  in terms of structural similarity index (SSIM), information leakage,   adversarial accuracy,  and reconstruction quality,  through   extensive  experimental studies in  different   communication scenarios. 
We also conduct \emph{ablation studies}  to further study the effect of different hyperparameters of 
	the implemented DNNs and  the loss function  on the E2E performance.

Finally, 	we would like to highlight  the main differences between our paper and the prior work of \cite{AE-Deniz}. 
	1)‌ Our proposed scheme considers both  AWGN and fading channels,  
	while \cite{AE-Deniz} only considers 
	an AWGN channel.   
	2) Our model is designed for privacy-aware  transmission against multiple eavesdroppers,  with different eavesdropping strategies,   
	while \cite{AE-Deniz} simply considers  one  Eve. 
	3) Our neural architecture differs  from  \cite{AE-Deniz}.
	We do not simply  employ  ``fully-convolutional'' DNNs, but also utilize generalized normalization transformation (GDN) blocks,   and their  inverse counterpart (IGDN),  \cite{gdn} at the hidden layers.  
	This 
	has been  shown to be able to 
	provide  significant improvements  in capturing image statistics 
 of natural images. 
	4) \textcolor{black}{The model is trained using a mixed
loss function that combines pixel-wise reconstruction (via MSE) with
\emph{perceptual quality optimization} (via the SSIM metric)}, whereas  
   \cite{AE-Deniz} only  considers the pixel-wise metric of  peak signal-to-noise ratio (PSNR).  
Unlike PSNR, the SSIM metric  is a perception-based metric, which captures pixel inter-dependencies, and hence, the semantics of the source image. 
	
	\vspace{0mm}
	\subsection{Paper Organization and Notations}
	{The remainder of this paper} is organized as follows. In Section \ref{sec:System_Model}, our proposed system model together with the main assumptions and intuitions are  provided. 
	We also formulate our problem to characterize the privacy-funnel-like  framework. 
	In 	Section \ref{sec:Proposed_approach} we   propose  our learning-based E2E approach to solve our 
	problem in a data-driven manner.    
	In  Section \ref{sec:evaluation},  we investigate the performance of our  scheme  through  extensive numerical  experiments. We also compare our learning-based scheme with different benchmarks. 
	Finally, Section \ref{sec:Conclusion} concludes our paper. 

	{\textit{Notations:} We denote the transpose, the conjugate transpose, and $\ell^2$ norm of a vector by $(\cdot)^\mathsf{T}$, $(\cdot)^\dagger$, and $||\cdot||$, respectively. 
		Vectors are represented by bold lowercase letters. 
		The zero and the identity matrices are shown by $\bm 0$ and $\bm {I}$, respectively.  
		$\mathcal{CN}(\mu,\sigma^2)$ and  
		$\mathcal{G}(k,\theta)$  represent a complex random variable (RV) with mean $\mu$ and standard deviation  $\sigma$, 
		and gamma distribution with
		shape parameter $k$ and scale parameter $\theta$, respectively.  
		The expected value and the probability density function (pdf) 
		of RV ${X}$ are denoted by $\E[{X}]$ and $p_X(x)$, 
		respectively. 
		The mutual information between RVs ${X}$ and ${Y}$ is denoted by  $I({X};{Y})$, while the cross-entropy  of  distributions $p$ relative to distribution $q$ over a given set is defined as ${H}(p,q) \overset{\Delta}{=}-\mathbb{E}_p[\log q]$.  Moreover,  $\text{KL}(\cdot || \cdot)$ denotes the Kullback-Leibler (KL) divergence. 
	}

	\begin{figure}
		\vspace{0mm}
		\centering
		\includegraphics
		[width=2.95in,height=1.95in,trim={0.0in 0.0in 0in  0.0in},clip]{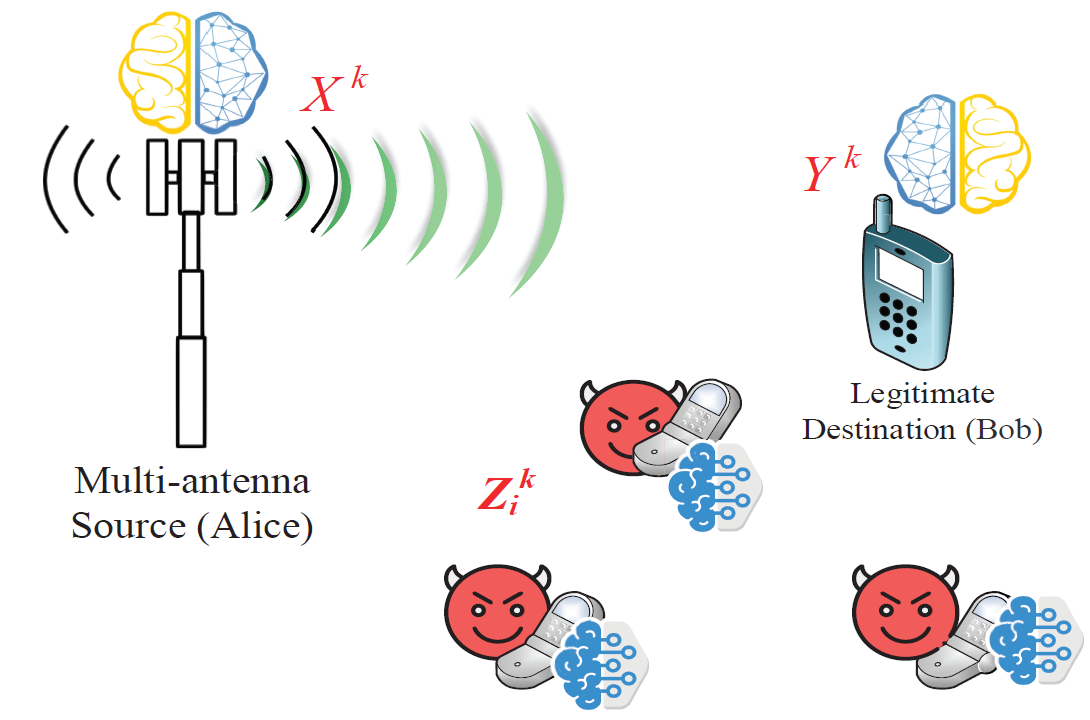}
		\vspace{-1mm}\caption{Proposed learning-based system model for end-to-end  image transmission against 
			multiple eavesdroppers.}
		\label{fig:SysModel}
  \vspace{-2mm}
	\end{figure}

	\vspace{0mm}
	\section{System Model and Problem Statement} \label{sec:System_Model}	 
    \subsection{Model Description} \label{System_Model} 
	Consider the communication scenario   demonstrated in Fig. \ref{fig:SysModel}, where a multi-antenna source  node, Alice ({$\cal A$}), aims to  transmit images {${U}^n \in \mathcal{U}^n$}  to a  destination  node, Bob ({$\cal B$}), 
	over $k$  uses of 
	the  wireless channel, 
  {where $\mathcal{U}$ denotes the alphabet of source images.}     
	According to the JSCC literature \cite{DJSCC-Deniz}, we refer to  the image dimension, $n$,  as the  {source bandwidth}.  The channel dimension $k$  characterizes  the {channel bandwidth}. We usually have  $k < n$,  which reflects the concept of {bandwidth compression}  (please see \cite{DJSCC-Deniz} and references therein for more details).  
	The image transmission service should be kept secret from multiple eavesdroppers, which overhear the communication  and are equipped with adversarial DNNs, aiming to extract private data regarding the source image. 
	$\cal A$ and $\cal B$ employ DNNs and perform \emph{secure  Depp-JSCC} by leveraging the concept of autoencoders. 
	Details of the 
	proposed DNN architectures, together with the training strategies of legitimate autoencoder  and adversarial neural decoders  are given in the next sections.

	Alice aims to convey the  source information $U^n$ to Bob  with minimum distortion, 
	while preventing the   information of  sensitive  part 
  {$S_i \in \mathcal{S}_i$, with a discrete alphabet $\mathcal{S}_i$,}  to be leaked to the $i$-th eavesdropper, $i = 1,\cdots,M$, where  $M$ denotes the number of eavesdroppers. 
	We note that the sensitive (private) parts are   correlated with $U^n$  with distribution  $p_{U^n,S_1, S_2 \cdots, S_M}$.  
	Toward this end, 
	Alice  maps the source information 
	$U^n$ into a {channel input codeword $X^k \in \mathcal{X}^k$} via implementing  a potentially 
	non-trivial  
	function {$f_{\cal A}: \mathcal{U}^n \rightarrow \mathcal{X}^k$}, where  $X^k = f_{\cal A}(U^n)$.  
	Transmitted codeword is subject to an average
	power constraint, $\frac{1}{k} \mathbb{E}[({X^k})^\ast{X^k}] \leq P$.
	Channel outputs at Bob and the $i$'th eavesdropper  are denoted, respectively, by $Y^k$ and $Z_i^k$, $i \in \{1, \cdots, M\}$.  
	Bob then applies a  decoding function
	$f_{\cal B}$ to obtain $\hat{U}^n = f_{\cal B}(Y^k)$. 
	Meanwhile, each eavesdropper tries to extract the sensitive information $S_i$,  that he is interested in, from his observations $Z_i^k$. 
	We consider the trade-off between  delivering images
	to Bob with the highest fidelity and  controlling  the information leakage to each adversary, which  
	is theoretically 
	measured	by the mutual information metric  $I({S_i};{ Z}_i^k)$. 
	In the case that all eavesdroppers are interested in the same secret we have  $S_i = S, \forall i$.

	Transmission of data-streams over the air  experiences  independent realizations of   communication channels.    
	Generally speaking, the channels impose random corruption on any  transmitted symbol vector ${\bm \tau} \in \mathbb{C}^k$, which 
	we model through
	the transfer function $\eta({\bm \tau})$. 
	We consider both the AWGN and  slow fading channels, where for the slow fading, we adopt two widely-used assumptions of Rayleigh fading and Nakagami-$m$ channels. 
	The transfer function of the Gaussian channel can be formulated by $\eta_{\nu} (\bm \tau)= \bm \tau + \bm \nu$, where the vector $\bm \nu \in \mathbb{C}^k$ is comprised of independent identically distributed (i.i.d.) samples from a circularly symmetric complex Gaussian distribution, i.e., $ \bm \nu \sim  \mathcal{CN}(\bm 0,\sigma^2\bm I_k)$, where $\sigma^2$ is the average power of additive noise.
	For the fading scenario, the multiplicative effect of the communication channel is modeled by $\eta_{h}(\bm z) =  h \bm \tau$, where $ h \sim \mathcal{CN}(0,\delta^2_h)$ for the Rayleigh fading scenario, and $|h|^2 \sim \mathcal{G}(m,\delta_h^2/m)$ for the Nakagami-$m$ scenario with $m>1$. 
	Here, $\delta_h$ 
	represents  the standard deviation (std) of the channel distribution, 
	which reflects the large-scale effects of channel fading.   
	The joint effect of channel fading and Gaussian noise can be treated by the composition of  transfer functions as 
	$\eta(\bm \tau) = \eta_\nu(\eta_h(\bm \tau))= h \bm \tau + \bm \nu$.

	Fundamentally,  the  trade-off between the minimum  achievable distortion at the  legitimate receiver and the minimum  leakage to Eve is asymptotically   characterized  in \cite{Yamamoto} under the idealistic assumption of i.i.d. source and channels with known distributions. 
	However, this result 
	is not  applicable  to practical systems 
	with finite block-lengths 
	and unknown source and channel statistics. 
	Hence, in our system model we consider a \emph{data-driven} approach for the  practical non-asymptotic regime. That is,  we do not consider  any underlying  assumption on the distribution of the data source  or the latent private variable.  
	{We remark that while we do assume a known channel model, we use this model to generate samples from conditional channel distribution. We could easily drop this assumption if we had data collected from a particular channel with unknown statistics.}

	\vspace{0mm}
	\subsection{Problem Formulation}\label{problem-form}
	\vspace{-1mm}
	Inspired by the key idea of \emph{privacy funnel} \cite{funnel},  we first formulate  an optimization   framework to characterize  the  trade-off between successful image recovery at Bob and the information leakage to the  eavesdroppers.  
	The formulated funnel-like framework is then  solved in a data-driven manner, utilizing DeepJSCC architecture. 
	That is, we implement  DNNs, 
	based on the concept of  autoencoders, to learn the   encoding-decoding functions, while 
	restraining the  adversarially-trained eavesdroppers from intercepting private information.  

	We aim to simultaneously minimize both the distortion $d(U^n,\hat{U}^n)$ at Bob and the information leakage,  $I({S_i};{ Z}_i^k), i \in [M]$,  about the secrets $S_i$. 
	This corresponds to a multi-objective programming (MOP) that can be solved using  the scalarization approach \cite{scale}. That is, we minimize a weighted sum  of the multiple objectives with  weights $w_i \in \mathbb{R}_{>0}$ as follows 
	\begin{align}\label{eq:P0}
		\underset{p_{X^k|U^n},
			f_{\cal B}}{\text{minimize}}
		\quad 
		\mathbb{E}\left[d(U^n,\hat{U}^n)\right] + 
		\frac{1}{M}	\hspace{0mm}  \sum_{i \in [M]} w_i I({S_i};{ Z}_i^k).  
	\end{align}
	Notably, $w_i$ in \eqref{eq:P0} 
    adjusts the trade-off between  the information leakages and the distortion at the legitimate receiver.    
	\textcolor{black}{To make \eqref{eq:P0} tractable, 
	we   further need to estimate the mutual information term.} In particular, it is still  a major challenge to optimize DNNs, while having an objective 
	function
	comprised of expressions about the mutual information between two (or more) data distributions.   
	It is also challenging to directly estimate the mutual information metric  from samples according to  \cite{Besser-FFNN,Eduard-AE,AE-Deniz, Ecenaz-icassp}.  
	\textcolor{black}{Hence,  an approximation of the mutual information metric is needed, in a way that it is also ``interpretable'' for the training of DNNs.} 
	
    \textcolor{black}{
In the following, 
inspired by \cite{AE-Deniz, Ecenaz-icassp},  we apply a \emph{variational 
	approximation}  
    for  the mutual information metric. First, we can write 
\begin{align}{\label{eq:MI_formula}}
    I(S_i;Z_i^k) &=  {H}(S_i) - {H}(s_i |  Z_i^k) \nonumber  \\ 
    & \overset{(a)}{=}  {H}(S_i) \hspace{-0.5mm}+\hspace{-0.5mm}  {\text{KL}}(p_{S_i|Z_i^k} || q_{S_i|Z_i^k})  \hspace{-0.5mm} + \hspace{-0.5mm} \mathbb{E}\hspace{-0.25mm}\left[\log q_{S_i|Z^{k}_{i}}(s_i|z_i)\hspace{-0.25mm}\right]  \nonumber \\ 
    & \overset{(b)}{=} 
    {H}(S_i) \hspace{-0.0mm}+\hspace{-0.25mm} 
    \underset{q_{S_i|Z^{k}_{i}}}{\text{max}}
    \mathbb{E}\hspace{-0.5mm}\left[\log q_{S_i|Z^{k}_{i}}(s_i|z_i)\hspace{-0.25mm}\right], 
\end{align}
where $(a)$ follows from the fact that $- {H}(s_i |  Z_i^k)  = {\text{KL}}(p_{S_i|Z_i^k} || q_{S_i|Z_i^k})   +   \mathbb{E} \left[\log q_{S_i|Z^{k}_{i}}(s_i|z_i) \right] $, and $(b)$ follows from the fact that KL divergence is nonnegative, where  the maximum is attained when the variational posterior $q_{S_i|Z_i^k}$ equals the true posterior $p_{S_i|Z_i^k}$,  i.e.,  $q_{S_i|Z_i^k} = p_{S_i|Z_i^k}$.  Due to the intractability  of the true posterior distribution,  $q_{S_i|Z^k_i}(s_i|z_i)$ can be characterized as the   likelihood estimation of the $i$'th adversary’s with respect to the correct	distribution of $S_i$, given the observation $Z^k_i$, which will be realized via adversarial DNNs employed at eavesdroppers.  Therefore, dropping the constant term $H(S_i)$ in \eqref{eq:MI_formula}, our objective function can be rewritten as   
    }  
    
	\begin{align}\label{eq:P1}
		\underset{p_{X^k|U^n},
			f_{\cal B}}{\text{minimize}} \quad \hspace{-2mm}\mathbb{E}\hspace{-1mm}\left[\hspace{-0.3mm}d(U^n,\hat{U}^n)\hspace{-0.3mm}\right]+\hspace{-2.5mm}‌\frac{1}{M}	\hspace{-1.5mm} \sum_{i \in [M]}  \hspace{-1.5mm}
		w_i 
		\underset{q_{S_i|Z_i^k}}{\text{max}}\hspace{-1mm}
		\mathbb{E}\hspace{-1mm}\left[\log q_{S_i|Z^{k}_{i}}(s_i|z_i)\hspace{-0.5mm}\right].  
	\end{align}
	\textcolor{black}{The benefit of this approximation is that we can  interpret the variational   lower  bound on information leakage  as the sample-wise cross-entropy  between the distribution over  adversaries' predictions, and the true 
	distribution of private attributes \cite{MI-approx, IT_CLUB}.}  
	
	Invoking \eqref{eq:P1}, one can infer  that 
	it can be viewed as a minimax game between Alice-Bob pair, and the set of adversaries.
	To elaborate,  while adversarial nodes  wish to maximize their individual  information leakage 
	by choosing the posterior likelihood  distributions $q_{S_i|Z^k_i}$, Alice and Bob  should jointly determine the optimum  encoding-decoding functions, $f_{\cal A}$ and $f_{\cal B}$, 
    that  minimizes  the  weighted sum of the distortion and leakage. 
	Nevertheless, we emphasize that  we do not  assume to have  any  knowledge about the underlying distribution of  image source and the sensitive attributes, which is aligned with the real-world scenarios. Our  scheme relies upon a \emph{data-driven} approach, i.e.,   
	we 
	leverage  datasets to  
	``learn'' the optimized privacy-aware  encoding-decoding functionalities.
	
	To solve the proposed MOP of \eqref{eq:P1} in a data-driven manner, we  implement  DNNs to parameterize the optimized   encoding-decoding functionalities  at Alice and Bob, while the adversaries also employ optimized adversarially-trained DNNs and try  to infer the sensitive attribute of the transmitted image.   
	Let $\Omega_{\cal A}$ and $\Omega_{\cal B}$  stand for the set of (trainable) parameters of the autoencoder pair at Alice and Bob, respectively. 
	In addition, let $\Theta_{E,i}$  parameterize   the adversarial network of the $i$'th adversary. 
	We can formulate the following  loss function to be minimized.   
	\begin{align}\label{eq:P-DL}
		\hspace{-2mm} \mathcal{L}(\Omega_{\cal A},\Omega_{\cal B},&\Theta_{E,1}, \cdots,\Theta_{E,M})     = 
	   \mathbb{E}\left[d(U^n,f_{\Omega_{\cal B}}(Y^k))\right]  \nonumber
		 \\ 
		& + 
		\frac{1}{M}	\hspace{0mm} \sum_{i \in [M]} w_i \hspace{2mm} \underset{\Theta_{E,i}}{\text{max}}
		\hspace{1mm}
		\mathbb{E}\left[\log q_{\Theta_{E,i}}({s_i|z_i})\right],  
	\end{align}
	where  $f_{\Omega_{\cal A}}$ and $f_{\Omega_{\cal B}}$  represent  the encoder and decoder functionality of Alice's DNN and Bob's DNN, respectively,   for which we have   $X^k = f_{\Omega_{\cal A}}(U^n)$. 
	Moreover, $q_{\Theta_{E,i}}({s_i|z_i})$ formulates the approximated adversarial   likelihood about the correct value $S_i=s_i$, 
	estimated  at the DNN of the  $i$'th adversary with parameters $\Theta_{E,i}$ and based on the observation $Z_i^k = z_i$.   
	We emphasize that similarly to  \cite{DJSCC-Deniz,Deniz-GDN,AE-Deniz},  the  legitimate  and wiretap   channels are 
	treated as non-trainable layers. 
	However, since  the considered channels are differentiable, we  incorporate  them
	as part of our proposed E2E  transmission scheme.

	In the following, we leverage  the novel  concept  utilized in generative adversarial networks (GANs) and iteratively train the DNNs   based on the loss function proposed in \eqref{eq:P-DL}.  
	To elaborate, each  joint training phase  of the autoencoder pair $(\Omega_{\cal A},\Omega_{\cal B})$ is  followed by a training step  for the adversaries' parameters $(\Theta_{E,1}, \Theta_{E,2},\cdots,\Theta_{E,M})$, where more details are given in the next section. 
	In what follows, we also 
	address  
	context-aware  interpretations for the distortion measure 
	and  the 
	likelihood distribution of sensitive attributes,  which  leads  us  toward   a  machine learning-interpretable approach for E2E privacy-aware  communication. 
	
	\begin{figure*}
		\vspace{0mm}
		\centering
		\includegraphics
		[width=6.0in,height=2.0in,trim={0 0.0in 0 0.0in},clip]{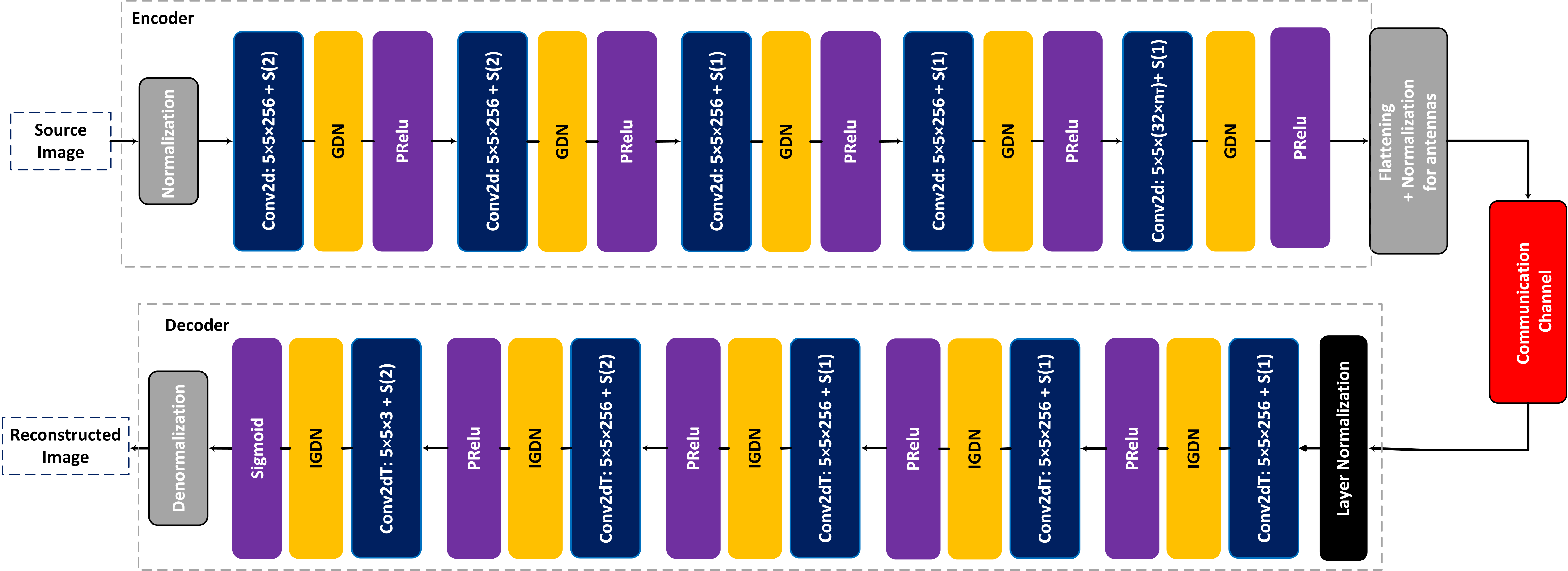}
		\vspace{0mm}
		\caption{Proposed deep neural networks at Alice (encoder) and Bob (decoder). 
			The notation $w\times w \times f$  denotes a convolutional layer with $f$ filters (channel outputs) of spatial extent 
			$w$. 
			The notation $w\times w \times (f \times n_{\sf T})$ at the last convolutional layer of encoder indicates that we require $f$
			filters for each of  $n_{\sf T}$ antennas at the  encoder output.  
			Moreover,  $s(\cdot)$ denotes  the stride, which  can be downsampling (at the encoder) or upsampling (at the decoder).}
		\label{fig:DNN}
	\end{figure*}

	\vspace{0mm}
	\section{Proposed Deep JSCC-Based Solution}\label{sec:Proposed_approach}
\begin{figure}
		\vspace{0mm}
		\centering
		\includegraphics
		[width=2.7in,height=1.45in,
		trim={0 0.0in 0 0.0in},clip]{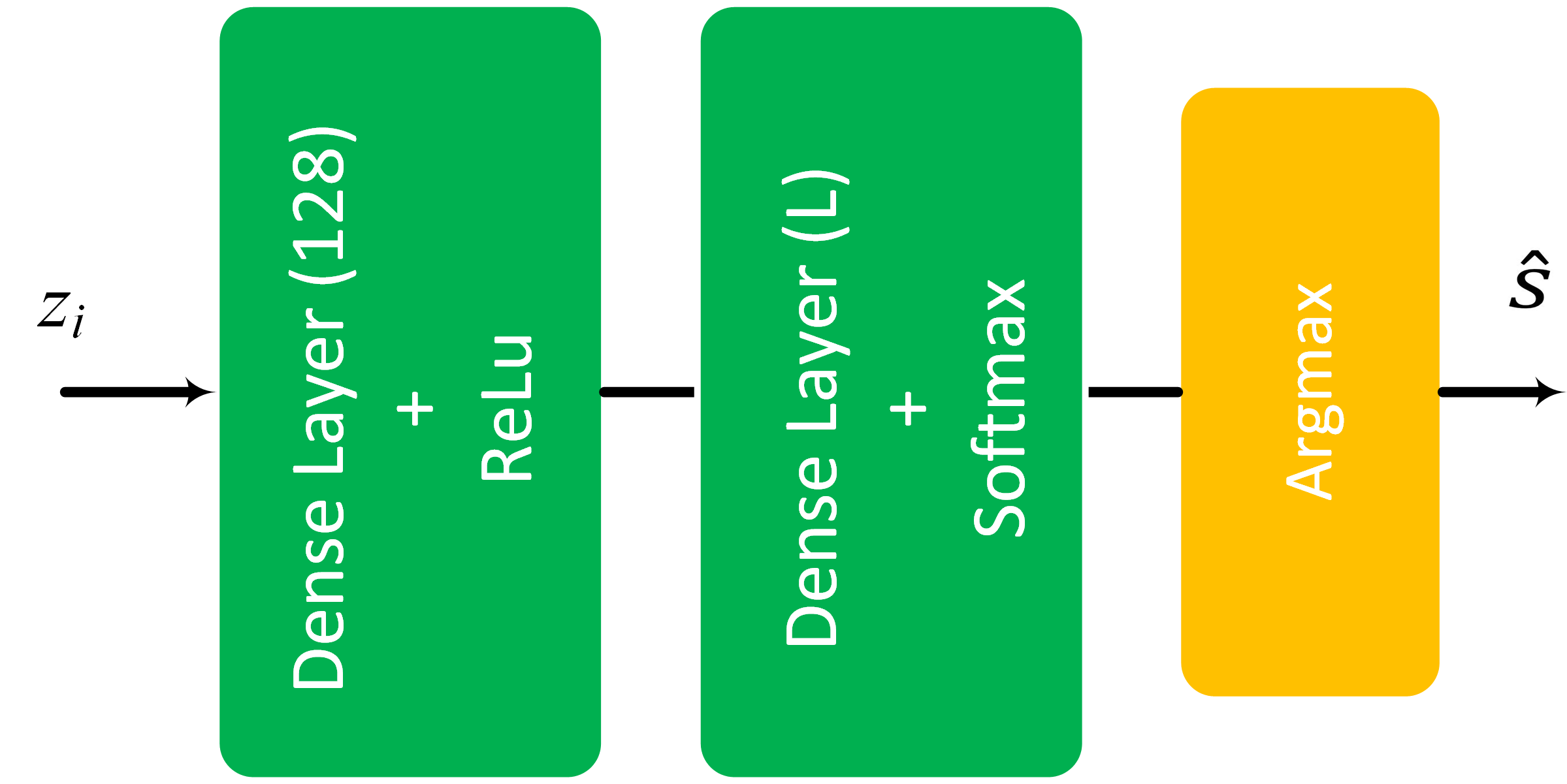}
		\vspace{0mm}
		\caption{Implemented DNN at each of the adversaries for extracting sensitive part of image files.}
		\label{fig:DNN-Advs}
  \vspace{0mm}
	\end{figure}

	Following the DeepJSCC concept, 
	we employ  DNNs  to  directly map the  pixel values to  the complex-valued samples,  which are  sent  over the air.  
	Consider image input files with dimensions $n = H \times W \times C$, where $H$, $W$, and  $C$ stand for the image height, width,  and  the number of channels ($3$ for colored images and $1$ for grayscale), respectively. 
	Alice maps  each realization of  the source data  $U^n$,  denoted by  $\bm  u \in \mathbb{R}^n$,   to a vector of   channel input   ${\bm x} \in \mathbb{C}^k$, which can be viewed as a realization of  $X^k$.  
	%
	${\bm x}$ should be securely encoded at Alice and decoded by Bob 
    based on the objective introduced in \eqref{eq:P-DL}.  
	This is done via our proposed DNN-based solution,   
	where the block diagram of the DNNs employed for  the neural encoder and decoder components of  legitimate parties  is illustrated in Fig. \ref{fig:DNN}. 
	In addition, the DNN employed by each of the  adversaries is demonstrated in Fig. \ref{fig:DNN-Advs}.  
	The  structure  of each  DNN component 
	is described next.\footnote{We state that the  considered  hyper-parameters for the  legitimate DNNs, as demonstrated in Fig. \ref{fig:DNN}, are  selected  based on  comprehensive  experiments and numerous trials, while the
		general architecture is inspired by \cite{Deniz-GDN}.}  
		
\vspace{0mm}


	\vspace{0mm}
	\subsection{Legitimate Neural Encoder-Decoder Pair:}
	At the legitimate encoder, we implement   convolutional layers, followed by  normalization invoked  by the generalized normalization transformations (GDN) block \cite{gdn}, which is then followed by a parametric ReLU (PReLU) \cite{PReLU} activation function.  
	At the output of the last 
	PReLU layer, which consisting of $2k \times n_{\sf T}$ elements, 
	we employ a flattening layer for each of the antennas,  to reshape the  encoded tensor  to a data-stream  
	from the latent space, resulting in a data sequence of the form  $\tilde{\bm x_i} \in \mathbb{C}^k$ for each antenna $1\leq i \leq n_{\sf T}$.    
	The encoded latent sequence is further  normalized, such that the channel input ${\bm X} \treq [\bm x_1, \bm x_2, \cdots, \bm x_{n_{\sf T}}] \in \mathbb{C}^{k \times n_{\sf T}}$ satisfies the average transmit power constraint $P$. Mathematically speaking, we have 

 \vspace{0mm}
	\begin{equation}
		{\bm x}_i = \sqrt{kP}\frac{\tilde{\bm x}_i}{\sqrt{\tilde{\bm{x}}_i^{\dagger}\tilde{\bm{x}}_i}}, \quad 1\leq i \leq n_{\sf T}. 
		\label{eq:power_constraint}
	\end{equation}
	Each column of the channel input data  matrix $\bm X$ is then 
	distributed over each of $n_{\sf T}$ antennas to exploit the available degrees of freedom.   

	The   joint source-channel coded output of Alice's DNN is   sent  over the communication channel, where  
	the  quality of the information-bearing sequence will be corrupted due to additive  noise, fading, or other channel impairments.  
	The  channel distorted version ${\bm y} \in \mathbb{C}^k$ 
	is observed by Bob, where ${\bm y}$ is considered  as a realization of $Y^k$, i.e., the   output of the communication channel. 
	Accordingly,  ${\bm y}$ is  fed to   Bob's DNN, which tries to recover the  image  by estimating  $\hat{\bm x} \in \mathbb{R}^n$ from
	${\bm y}$, 
	resulting in an approximate reconstructed version  of the original data. 
	Specifically, the real and imaginary parts of the $k$ (complex-valued) channel output samples 
	form $2k$ real-valued elements,  
	which are fed into the
	convolutional layers.      
	Based on the encoding conducted by Alice,  Bob's DNN   inverts the operations implemented  at 
	Alice by passing the observed  data-stream  ${\bm y}$ through a series  of transposed convolutional layers  followed by the inverse of  GDN (IGDN) blocks and  PReLU activation functions.  
	
    \vspace{-0.5mm}
	The intuition behind the  proposed  architecture  
	 can be explained as follows. 
	Convolutional layers are capable of  extracting  image features. 
	After that the  convolutional layers have  extracted the features, the GDNs (which perform differentiable and invertible operations) conduct  local divisive normalization, which has been shown to be effective  in  density modeling of data \cite{gdn}.  
	Moreover, the GDNs offer  significant  improvements  in capturing image statistics  through Gaussianizing the data of natural images.  
	In the next step, the 
	activation functions  facilitate the learning process  of non-linear mappings  from the source image  space to  the latent space, i.e., the channel input, and vice versa.  
	Finally, the hyperparameters of Bob's DNN mirror the corresponding functionalities performed by the encoder layers of  Alice's DNN.  
	
	{\textit{Remark 1:‌}}
	We  remark that before running the proposed DNNs, the input images are normalized by the maximum pixel value, e.g.   $255$ in this paper, to produce values in the range of $[0,1]$. This operation will be  inverted at the decoder (as demonstrated in Fig. \ref{fig:DNN}) to reconstruct (de-normalized) pixel values within  $[0, 255]$ range.  
	The rationale  behind this reformulation is that the dependence of  DNNs on the  maximum pixel values is relaxed, since the statistics of data are unknown at the decoder. 
	This also facilitates the training process  of our proposed DNNs \cite{arxive}.  
	Notably, the sigmoid activation function applied to the output of Bob's decoder  is  for producing normalized approximation ($\hat{\bm{x}}$) for the original data  in the  range of $[0,1]$ before de-normalizing. 
	
	{\textit{Remark 2:‌}}
	Prior to performing DNN-based  decoding operations, a layer normalization (LN) block \cite{norm-layer} is implemented at Bob's DNN.
	The LN block is aimed to  realize  
	data normalization 
	via taking into account   
	all of the summed inputs to the 
	Bob's DNN 
	during \emph{every single training or inference sample}. 
	Notably,  unlike batch normalization  which  considers  the distribution of the summed input to a neuron over \emph{mini-batches} of training set, LN performs  the same computation at training and inference phases
	\emph{per data sample}. 
	It is also shown in \cite{norm-layer} that compared with previous techniques,  LN substantially reduces the training time.

	\vspace{-1mm}
	\subsection{Adversarial Neural Networks for Eavesdroppers}
    \vspace{-1mm}
	The structure of  adversarial DNNs, which are parameterized by $\Theta_{E,i}, i \in [M]$,   is presented  in Fig. \ref{fig:DNN-Advs}.
	According to the proposed system model in Section \ref{System_Model}, the adversaries 
	utilize DNNs to facilitate the extraction  of  sensitive information $S_i$ from their received signals. The sensitive information $S_i$ can be assumed to be the class to which the images belong \cite{AE-Deniz, Ecenaz-icassp}.    
	For instance, the identity of patients within  medical imaging in e-health applications,  or 
	the locations of critical infrastructures in a cyber-physical system.    
	To 
	extract sensitive information (e.g.,  privatized data) from images,  each adversary employs the DNN architecture illustrated in Fig. \ref{fig:DNN-Advs} as a class predictor, which is  comprised of   
	a dense layer (fully-connected network with $128$ neurons) with rectified linear unit (ReLU) activation function.\footnote{\textcolor{black}{While
our evaluation focused on specific DNN architectures, the framework is model-agnostic
and can incorporate more advanced neural architectures, such as transformers. Model-agnostic theoretical bounds on information leakage against advanced architectures are left for future works.}} The dense layer is  concatenated  by another dense layer with softmax activation, where the  dimension of the  output  neurons, $L$, equals the 
	cardinality  of the secret $|S_i|$. 
	The output of the softmax layer produces  an adversarial likelihood  estimation regarding the posterior  distribution  $q_{\Theta_{E,i}}({s_i|z_i})$  of  sensitive attribute.\footnote{An argmax layer is added over the obtained distribution to take a guess about the predicted secret class $\hat{s}$. This also helps  assess the performance of  adversaries  via  measuring their accuracy 
		as the fraction of correct guesses. 
	} 
	Specifically, considering the output ${\bm z_i} \in \mathbb{C}^k$ of the $i$'th wiretap channel  (a realization of $Z_i^k$) as the observation of the $i$'th eavesdropper, 
	it is	  fed to the  adversarial DNN to obtain an approximated prediction $q_{\Theta_{E,i}}({s_i|z_i})$ regarding the sensitive attribute. 
	
	Invoking \eqref{eq:P-DL}, each adversary  tries to minimize its cross-entropy between the adversarially-estimated likelihood distribution and the ground-truth, where the ground-truth is represented by the one-hot encoded vector of $S_i$, as $ {{\bm \varepsilon}_s}_i \in \{0,1\}^L$.   
	Having lower cross-entropy values results in  
	higher similarity between the posterior adversarial likelihood 
	$q_{\Theta_{E,i}}({s_i|z_i})$
	and the ground-truth ${{\bm \varepsilon}_s}_i$, which can be interpreted as increase in the information leakage. 
	Meanwhile, Alice and Bob  jointly try to  minimize the reconstruction distortion $d(\cdot, \cdot)$ and the information leakage (approximated by the negative cross-entropy). 
	Exploiting the above-mentioned  learning-based interpretation,  
	our  sample-wise  framework  can be reformulated as  
	\begin{align}\label{eq:P-DL-CE}
		\text{minimize}	\hspace{2mm} & \mathcal{L}(\Omega_{\cal A},\Omega_{\cal B},\Theta_{E,1}, \Theta_{E,2},\cdots,\Theta_{E,M})  = 
		\nonumber \\ & 
		\mathbb{E}_{p(\bm{u},\bm{\hat{u}})} \left[	d\left(\bm u,f_{\Omega_{\cal B}}(\bm y)\right)
		\right] 
		\nonumber \\ 
		& + 
		\frac{1}{M}	\hspace{-1mm} \sum_{i \in [M]}  w_i \hspace{1mm}
		\underset{\Theta_{E,i}}{\text{max}}
		\hspace{0.5mm}
		\left(- H\hspace{-1mm}\left(q_{\Theta_{E,i}}({s_i|z_i}),{{\bm \varepsilon}_s}_i\right)\right), 
	\end{align}
	where $\hat{\bm u} = f_{\Omega_{\cal B}}(\bm y)$, and 
	$p(\bm u, \hat{\bm u})$ stands for the joint probability distribution of the original and the reconstructed image. 

\textcolor{black}{\emph{Remark 3:} 
The choice of cross-entropy as the privacy leakage metric has some  advantages in terms of computational efficiency,  interpretability, and more stable training. Notably, cross-entropy offers several practical advantages in scalable neural network-based environments, compared to other  privacy measure such as mutual information and differential privacy (DP). Cross-entropy naturally integrates into DNN training as a stable, differentiable loss function, unlike mutual information  which cannot be optimized directly, whilst mutual information is often approximated from cross-entropy to enable empirical calculation in the DNN training process. Moreover, a small DP noise can cause unpredictable utility loss, which can de-stabilize  training.}
    
	Due to the fact that  the true distribution of data $p(\bm u)$ is often unknown, obtaining  an analytical (closed-form) solution for  \eqref{eq:P-DL-CE} is not tractable \cite{AE-Deniz,DJSCC-Deniz,Deniz-GDN,Eduard-AE}. 
	Thus,  we estimate the expected distortion measure  using samples ${\bm u}_j$ of available  datasets by computing $\mathbb{E}_{p(\bm{u},\bm{\hat{u}})} \left[	d\left(\bm u,f_{\Omega_{\cal B}}(\bm y)\right)
	\right] \approx \frac{1}{N_{\cal T}}\sum_{j=1}^{N_{\cal T}} d(\bm u_j, \hat{\bm u}_j)$, where   ${N_{\cal T}}$ stands for  the number of training samples. 	
	{We note that, in this framework, we are assuming that we know the sensitive attribute each eavesdropper is interested in and their channel models, both of which are common assumptions in the privacy \cite{AE-Deniz, Ecenaz-icassp, funnel} and wiretap channel \cite{Eduard-AE, Yamamoto} literature.}   
	

	\vspace{0mm}
	\subsection{Training Procedure}\label{sec:training}
	In order to train our   learning-based approach based on  the objective  function proposed in \eqref{eq:P-DL-CE}, we  follow an  iterative procedure.  
	Inspired by the key idea of GANs, 
	each  joint training phase  of the autoencoder pair $(\Omega_{\cal A},\Omega_{\cal B})$ is  followed by a training step  for the 
	adversarial networks 
	$(\Theta_{E,1}, \Theta_{E,2},\cdots,\Theta_{E,M})$.  
	Intuitively, we are faced  with a  minimax game, i.e., the competition between legitimate autoencoder and the adversarial  DNNs.  
	Therefore,  the following strategy is run  through our proposed E2E system. 	
	
	\begin{itemize}
		\item[(i)]
	 Alice and Bob try to jointly minimize their loss function, 
        \begin{align}
            \hspace{-5.5mm} \mathcal{L}_{\cal AB} \hspace{-1mm} = \hspace{-1mm} \frac{1}{N_{\cal T}}\hspace{-1mm}\sum_{j=1}^{N_{\cal T}}
			\hspace{-1.5mm} \left( \hspace{-1mm}
			d(\bm u_j, \hat{\bm u_j})\hspace{-1mm}-
			\hspace{-1mm}\frac{1}{M} \hspace{-2mm}\sum_{i \in [M]} \hspace{-2mm} w_i \hspace{0mm}
		H\hspace{-1.5mm}\left(\hspace{-0.5mm}q_{\Theta_{E,i}}({s^{(j)}_i|z^{(j)}_i}),{{\bm \varepsilon}_{s}}^{(j)}_i \hspace{-0.5mm} \right)\hspace{-1.5mm}
			\right)\label{eq:L_AB}
        \end{align}

		
		To reinforce system's security against adversaries, we take one further step within  the training process of legitimate nodes. We perform \emph{adversarial likelihood compensation} (ALC), which has been shown to be more effective in confusing an adversary than the data-dependent one-hot encoding approach \cite{AE-Deniz}.   
		The main idea behind 
		 ALC is that we want to make the estimated likelihood of  adversaries 
		imitate a uniform distribution $\bar{p}_L = [\frac{1}{L},\cdots,\frac{1}{L}]^{\sf T}$. 
		By doing so, {Alice and Bob try to jointly maximize the uncertainty of adversarial  predictions
			by maximizing the similarity (minimizing the cross-entropy) between the adversarial likelihood and the uniform distribution $\bar{p}_L$,}  
		instead of  minimizing the likelihood corresponding to the correct prediction ${{\bm \varepsilon}_s}_i$. 
		Hence, a revised  objective function for Alice-Bob pair is formulated as 
        \begin{align}
           \hspace{-5.5mm} \mathcal{L}^{\sf ALC}_{\cal AB} \hspace{-1mm} = \hspace{-1mm} \frac{1}{N_{\cal T}}\hspace{-1mm}
		\sum_{j=1}^{N_{\cal T}} \hspace{-1mm}
		\left( \hspace{-1mm}
		d({\bm u_j}, {\hat{\bm u}}_j) \hspace{-1mm}
		+  \hspace{-1mm}
		\frac{1}{M} \hspace{-1.5mm} \sum_{i \in [M]} \hspace{-1.5mm} w_i 
		H\hspace{-1.5mm}\left(q_{\Theta_{E,i}}({s^{(j)}_i|z^{(j)}_i}),\bar{p}_L \hspace{-0.5mm}\right)\hspace{-1.6mm}
		\right) \label{eq:L_AB-LE}
        \end{align}
		\item[(ii)] 
		Each step of the  joint training for $\cal A$-to-$\cal B$ autoencoder  $(\Omega_{\cal A},\Omega_{\cal B})$ is  followed by a training step  for the adversarial DNNs.
		Eavesdroppers seek to minimize the cross-entropy between their estimated likelihood $q_{\Theta_{E,i}}({s_i|z_i})$ and the one-hot  vector ${{\bm \varepsilon}_s}_i$ corresponding to $S_i$. 
		That is, we have the following loss function for training the adversarial networks.    
		\begin{align}\label{eq:L_E}
			\hspace{-4mm}{\cal L}_{E} =  \frac{1}{N_{\cal T}}\sum_{j=1}^{N_{\cal T}} 
			H\hspace{-1mm}\left(q_{\Theta_{E,i}}({s^{(j)}_i|z^{(j)}_i}),{{\bm \varepsilon}_{s}}^{(j)}_i\right), \quad \hspace{-2mm} i \in [M]
		\end{align}  
	\end{itemize}	
	Note that all the adversarial networks can be trained in parallel. 	
	The distortion measure we consider for our legitimate loss function $	\mathcal{L}_{\cal AB}$ is a mixture of the average MSE, denoted by ${\Delta}^{\sf{MSE}}$, and 
	SSIM, ${\Delta}^{\sf{SSIM}}$,  between the  input image $\bm u$  and the recovered pixels $\hat{\bm u}$  at the output of Bobs' DNN.
	Therefore, we assume $d(\cdot,\cdot)$ to be measured  as follows 
	\begin{align}\label{eq:distortion}
		d(\bm u , \hat{\bm u}) = 
		{\Delta}^{\sf{MSE}} + 
		\alpha {\Delta}^{\sf{SSIM}},
	\end{align}
	where 
	\begin{align}
		\Delta^{\sf MSE} & =\frac{1}{n}||\bm u-\hat{\bm u} ||^2,  
		\quad  \Delta^{\sf SSIM} = 1- 	\mathsf{SSIM}(\bm u, \hat{\bm u}),    
	\end{align}
	with $n$ denoting the size of the original image  vector $\bm u \in \mathbb{R}^n$. Moreover, $\alpha$ is a tuning parameter representing the contribution of SSIM metric in the distortion measure $d(\cdot,\cdot)$.  
	The SSIM measure between two images $I$ and  $K$ is defined as 
	\begin{align}\label{eq: ssim}
		\mathsf{SSIM}(I, K) \overset{\Delta}{=} \frac{(2\mu_{I}\mu_{K}+c_1)
			(2\sigma_{IK}+c_2)}{(\mu^2_{I}+\mu^2_{K}+c_1)(\sigma^2_{I}+\sigma^2_{K}+c_2)}, 
	\end{align} 
	where   $\mu_{I}$, $\mu_{K}$, $\sigma_{I}$, $\sigma_{K}$, and $\sigma_{IK}$ are the local means, standard deviations, and cross-covariance for images $I$ and  $K$. $c_1$ and $c_2$ are two adjustable constants \cite{ssim-learning-2}.
	We note that the SSIM index is a  \emph{perception-based metric}, which is  able to capture  pixel inter-dependencies even for spatially close pixels.   
	%
	The rationale behind assuming
	the  proposed distortion metric  in \eqref{eq:distortion} 
	is that we not only aim to recover every pixel of a source image with minimum error (realized via MSE measure), but also want to obtain a 
	good-quality  reconstruction from the human visual system point of view.  
	We also remark that SSIM calculation is relatively simple, and its derivatives can be   easily computed within the gradient descent-based algorithms  \cite{ssim-learning}. 
    Training procedure can be carried out using off-the-shelf stochastic gradient descent algorithms such as the widely-adopted  adaptive moment estimation (Adam) optimizer \cite{adam}.   

\textcolor{black}{
\emph{Remark 4:} 
The main bottleneck for memory and optimization of the proposed framework lies in step (i) of the training procedure. This is due to the fact that  all neural networks (Alice, Bob, and all Eves) are involved in training, and the entire computational graph 
needs to be updated through backpropagation. To mitigate the computational bottleneck of joint training in case the number of adversaries is large, one can define  a maximum number of Eves to participate in each step of  backpropagation.  This can be modeled as a scheduling problem, based on memory storage constraints.  
For example, if a single adversary can be chosen at each step, each batch of image samples is  passed through Alice, Bob, and the chosen  Eve every time, realizing ``sequential adversarial training.'' 
}

	\vspace{-1mm}
	\begin{table}
		\centering
		\small
		\caption{\small Parameters for Training the Proposed System}\label{Tab1}
		\vspace{-1mm}
		\begin{tabular} 
			{|p{2.85in}|p{0.55in}|}
			\hline \textbf{Learning Parameters} & \textbf{Values}\\
			\hline
			\hline
			Batch size (${m}$) & 128\\
			Maximum number of training episodes ($N_{\text{episode}}$) & 200\\
			Number of 
			warm-up epochs ($N_\text{warm-up}$) & 50\\
			Number of legitimate training epochs per episode  
			($N_{\text{L}}$) & 5\\
			Number of adversarial training epochs per episode 
			($N_{\text{E}}$) & 5\\
			Training SNR of legitimate link ($\Gamma^{\sf train}_{{L}}$) & 20\\ 
			Training SNR of adversarial links ($\Gamma^{\sf train}_{{E}}$) & 15\\
			Learning rate  & $10^{-4}$\\
			Learning rate drop factor & 0.9\\
			Optimizer & Adam \cite{adam}\\
			\hline
		\end{tabular} \vspace{0mm}	
	\vspace{0mm}
	\end{table}

\begin{figure}
    \centering
    \includegraphics[width=3.25in,height=2.5in,
			trim={0.05in 0.0in 0 0.0in},clip] 
			{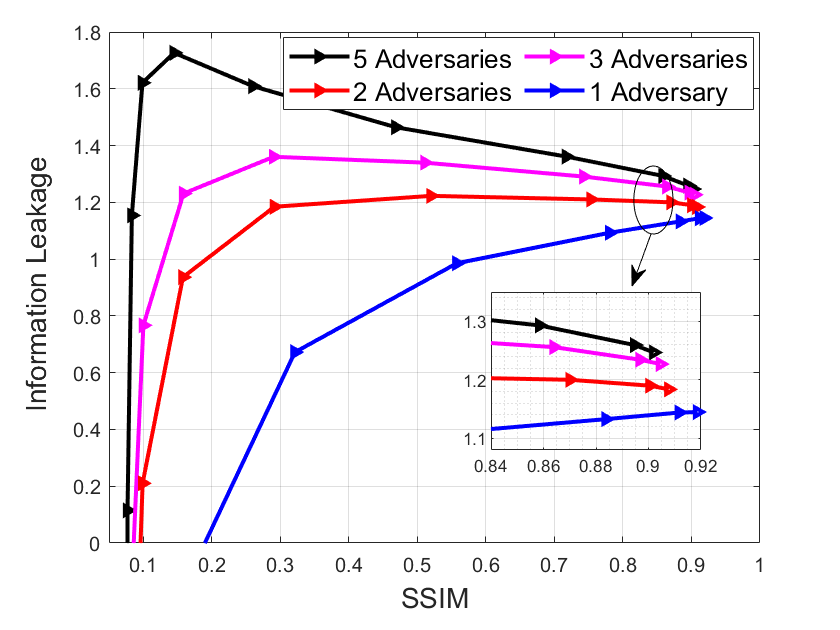} 
				\caption{\textcolor{black}{Privacy-utility trade-off over CIFAR-10. 
            }}\label{fig:util}
\end{figure}

 \begin{figure*}
		\centering
		\begin{minipage}[b]{0.5\textwidth}
			\includegraphics 
			[width=3.4in,height=2.85in,
			trim={0in 0in 0 0in},clip] 
			{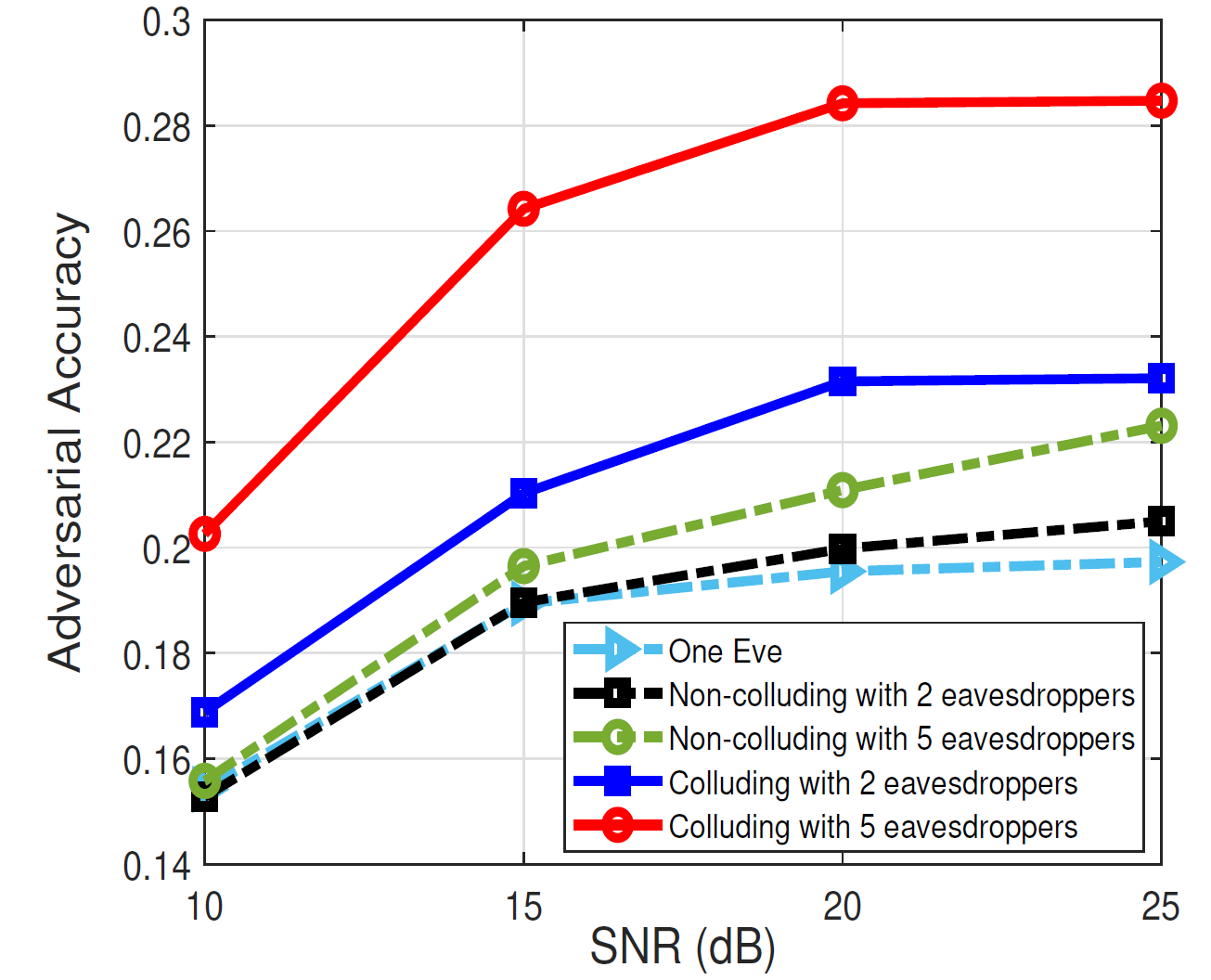} 
            \vspace{-1mm}
			\caption*{{(a) Adversarial accuracy (Rayleigh)}}
			\label{fig:acc}
		\end{minipage}\hfill
		\begin{minipage}[b]{0.5\textwidth}
			\includegraphics
			[width=3.4in,height=2.85in,
			trim={0.0in 0.0in 0 0.0in},clip]
			{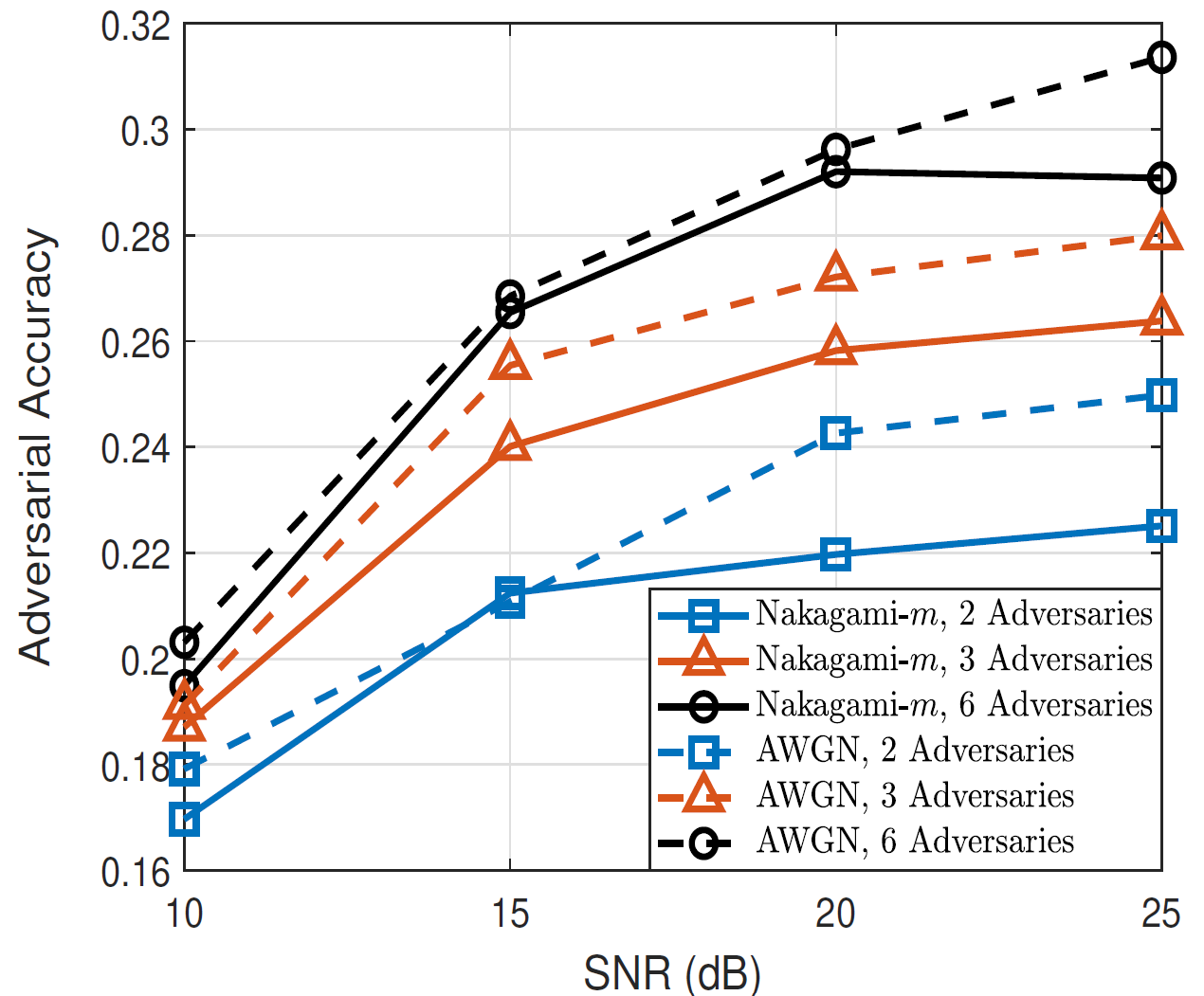}
            \vspace{-2mm}
			\caption*{(b) Adversarial accuracy (Nakagami-$m$)}
			\label{fig:acc_Nak}
		\end{minipage}
        \vspace{0mm}
		\caption{
		Adversarial accuracy vs. SNR ($\Gamma_{E}$) for Rayleigh fading, AWGN, and Nakagami-$m$  (with $m=3$) channels.}
		\label{fig:sut_and_acc}
		\vspace{-2mm}
	\end{figure*}

	\vspace{-1mm}
	\section{Implementation and Evaluation}\label{sec:evaluation}
	In this section, we present different experiments  to demonstrate the performance of  our proposed deep learning-based approach in various scenarios.   Different  benchmarks and ablation studies are also provided. In our experiments, 
	{we consider both scenarios of colluding and non-colluding eavesdroppers. 
		In addition, we study both cases of eavesdroppers being interested in a common secret $S$, and the case in which every eavesdropper is interested in a different secret $S_i$, using two different datasets of CIFAR-10 \cite{CIFARdataset} and CelebA \cite{CelebA}, respectively.}
	We examine the performance over  both  AWGN and complex  fading (Rayleigh and Nakagami-$m$) communication channels for  different channel SNRs and $4$ antennas (unless otherwise stated). 
	Moreover, we  address  the generalization capability of our proposed scheme for different communication scenarios and over a wide range of channel SNRs. 
	We also study the \emph{privacy-utility trade-off} for the proposed learning-based approach, which provides useful insights.    
\textcolor{black}{To quantify the level of privacy, we evaluate the information leakage characterized via the mutual information metric. Accordingly, with a similar approach to what proposed in \eqref{eq:MI_formula},  we use the following approximation for mutual information leakage   
\begin{align} 
   I(S_i, Z_i^k) \approx H(S_i) - H\hspace{-1mm}\left(q_{\Theta_{E,i}}({s_i|z_i}),{{\bm \varepsilon}_s}_i\right), 
\end{align} 
where the entropy of private attributes $H(S_i)$ are numerically calculated for each of the datasets according to the empirical distribution of CIFAR-10 and CelebA classes.    
}      
	
The SNRs of communication links are  defined as:
	\begin{equation}
		\Gamma_{L}=10\log_{10}\frac{P}{\sigma_{L}^2}~ \mathrm{~(dB)}, \quad \Gamma_{E}=10\log_{10}\frac{P}{\sigma_{E}^2}~ \mathrm{~(dB)}, 
		\label{eq:snr}
	\end{equation}
	which represent the ratio of the average power of the encoded data-stream of the latent space (i.e., the channel input) to the average noise power of legitimate $\sigma^2_{L}$ and adversarial nodes $\sigma^2_{E}$, respectively. 
	Without loss of generality, we set the average signal power  to $P=1$ for all experiments, while  varying  the SNR by setting the noise variances $\sigma_{L}^2$ and $\sigma_{E}^2$.  
The 
		std of the eavesdropping  channels are set to 
		$\{0.04, 0.16, 0.36, 0.64, 1, 1.44\}$  
         for  $i \in [6]$, for $M = 6$ simulated eavesdroppers in total.  
	The bandwidth compression ratio 
	is set to  $\frac{k}{n}=\frac{1}{3}$. 
	For the training, we have assumed the general case of complex Rayleigh  fading channel model which is treated by the joint effect of channel fading and additive  noise, i.e.,   
	$\eta(\bm \tau) = \eta_\nu(\eta_h(\bm \tau))= h \bm \tau + \bm \nu$ 
	with  $h \sim \mathcal{CN}(0,1)$. 
	Nevertheless, during the inference phase we study the performance of our proposed scheme in  different scenarios of AWGN and Nakagami-$m$ channels $|h|^2 \sim \mathcal{G}(m,\delta_h^2/m)$.

	For simplicity, we consider a single weight in the loss function, that is, $w_i=w, \forall i$.
	{We set $w = 5$ and $\alpha = 0.1$ during training.}   
    \textcolor{black}{The values of $w$ and $\alpha$ were chosen empirically.  We  first defined a set of plausible values for $\alpha$ and  $w$, trained the system for all combinations of $(\alpha, w)$ pairs, and evaluated the performance metrics on a validation set, and finally chose the best desired parameters.  We used TensorBoard\footnote{\url{https://www.tensorflow.org/tensorboard}} for tracking and visualizing performance metrics during training, and we ran {multiple combinations} of $\alpha$ and $w$, and  constantly monitored the loss function and other performance metrics such as SSIM, MSE, and cross-entropy over epochs. 
In Appendix \ref{app_TP}, the evolution of performance metrics over the training steps are visualized. 
Our observations have shown that for a given SNR, the performance is not too sensitive to the choice of these parameters, (although there is no guarantee that a selected set of parameters will be optimal for all ranges of SNRs).
}
    
	We have implemented our proposed network architecture (shown in Figs. \ref{fig:DNN} and \ref{fig:DNN-Advs}) using  {Python3 with  Tensorflow  \cite{tf}.}  
	{The codes  were run  
		on Intel(R) Xeon(R) Silver 4114  CPU running at 2.20 GHz with GeForce RTX 2080 Ti GPU.}
	To minimize the legitimate and adversarial loss functions (LF), 
	we have chosen the widely-adopted Adam optimizer algorithm   \cite{adam}. 
		Similarly to \cite{AE-Deniz}, we  choose  to fix the number of training episodes ${N}_{\text{episode}}$ in advance, where the value of ${N}_{\text{episode}} = 200$ is set according to our  experiments.
	The learning parameters used during the training process are summarized in Table \ref{Tab1}.       
	The detailed description of training process can be stated as follows. 
	i) During a \textbf{\emph{warm-up phase}}, both legitimate pairs and adversarial networks are independently trained in order to reach  a plausible initial state of reconstructing the image.  To elaborate, for $N_{\text{warm-up}} =50$ epochs,  the encoder-decoder pair of Alice and Bob $(\Omega_{\cal A}, \Omega_{\cal B})$ are trained according to the reconstruction  objective of minimizing $\mathcal{L}^{\text{warm-up}}_{\cal AB} =  \frac{1}{N_{\cal T}}\sum_{i=1}^{N_{\cal T}} d(\bm u_i, \hat{\bm u_i})$, where $d(\cdot, \cdot)$ is given in \eqref{eq:distortion}. Following the legitimate warm-up,  the adversarial networks $(\Theta_{E,1},\cdots,\Theta_{E,M})$ are trained for $N_{\text{warm-up}} =50$ epochs, while the legitimate encoder $\Omega_{\cal A}$ is frozen. 
	ii) After performing  the warm-up, the main phase of \textbf{\emph{minimax training}}   starts. In this phase, for each of the $N_{\text{episode}} = 200$ episodes,  Alice and Bob aim to learn the optimal mapping for the latent space and the decoding neural structure so that the reconstruction distortion and information leakage are minimized.
	Hence,  the legitimate encoder-decoder counterparts $(\Omega_{\cal A}, \Omega_{\cal B})$ are trained for $N_{\text{L}} = 5$ epochs based on minimizing the LF $\mathcal{L}_{\cal AB}^{\sf ALC}$ given in \eqref{eq:L_AB-LE}.   
	Since the neural encoder $\Omega_{\cal A}$ at Alice has learned a new  embedding for the original data, the adversaries are then trained for $N_{\text{E}} = 5$ epochs to 
	learn and adjust their adversarial role in extracting sensitive information from the newly-learned latent space based 	on the LF, ${\cal L}_{E}$, given in \eqref{eq:L_E}. 
	In this manner, the concept of minimax game between $(\cal A, \cal B)$ and the set of eavesdroppers   is realized, while maintaining the  fairness 
	for  the 
	learning process of legitimate  and adversarial nodes.  


	\vspace{-2mm}

	\subsection{Evaluation on CIFAR-10 dataset}
	\label{sec:cifar10_eval}
	We evaluate our proposed  framework using images with dimension 32 $\times$ 32 $\times$ 3 (height, width, channels) from the CIFAR-10 dataset \cite{CIFARdataset}. 
	The dataset  consists of $60000$  colored images of size $32\times32$ pixels. The training and evaluation sets are two  completely separated sets of images,  containing  50000 and 10000 images, respectively,  
	associated with  $10$ classes.  
	Adversaries wish to 
	infer a common secret $S$ 
	from the received noisy encoded signals. 
	Hence, the common secret $S$ here is considered as the  class of images with $|S| = L = 10$.

	\begin{figure*}
		\centering
		\begin{minipage}{0.45\textwidth}
			\centering
			\includegraphics
			[width=3.4in,height=2.5in,
			trim={0in 0.0in 0 0.0in},clip] 
			{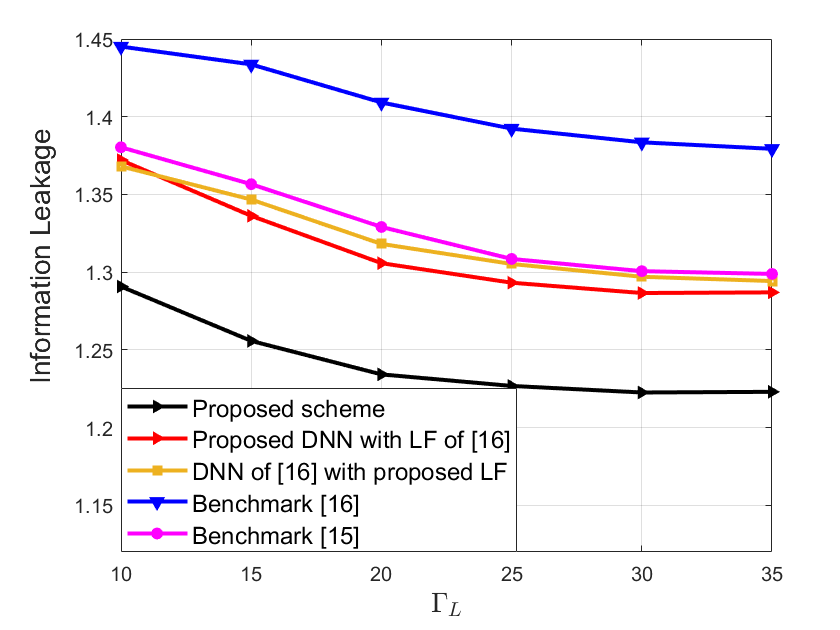}
            \vspace{-4 mm}
			\caption*{\textcolor{black}{(a) Privacy performance in terms of information leakage $I(S; Z_i^k)$ }}
			\label{fig: ce_ablation_test}
		\end{minipage}\hfill
		\hspace{0mm}
		\begin{minipage}{0.45\textwidth}
				\centering
			\includegraphics
			[width=3.4in,height=2.5in, 
			trim={0.0in 0.0in 0 0.0in},clip]
			{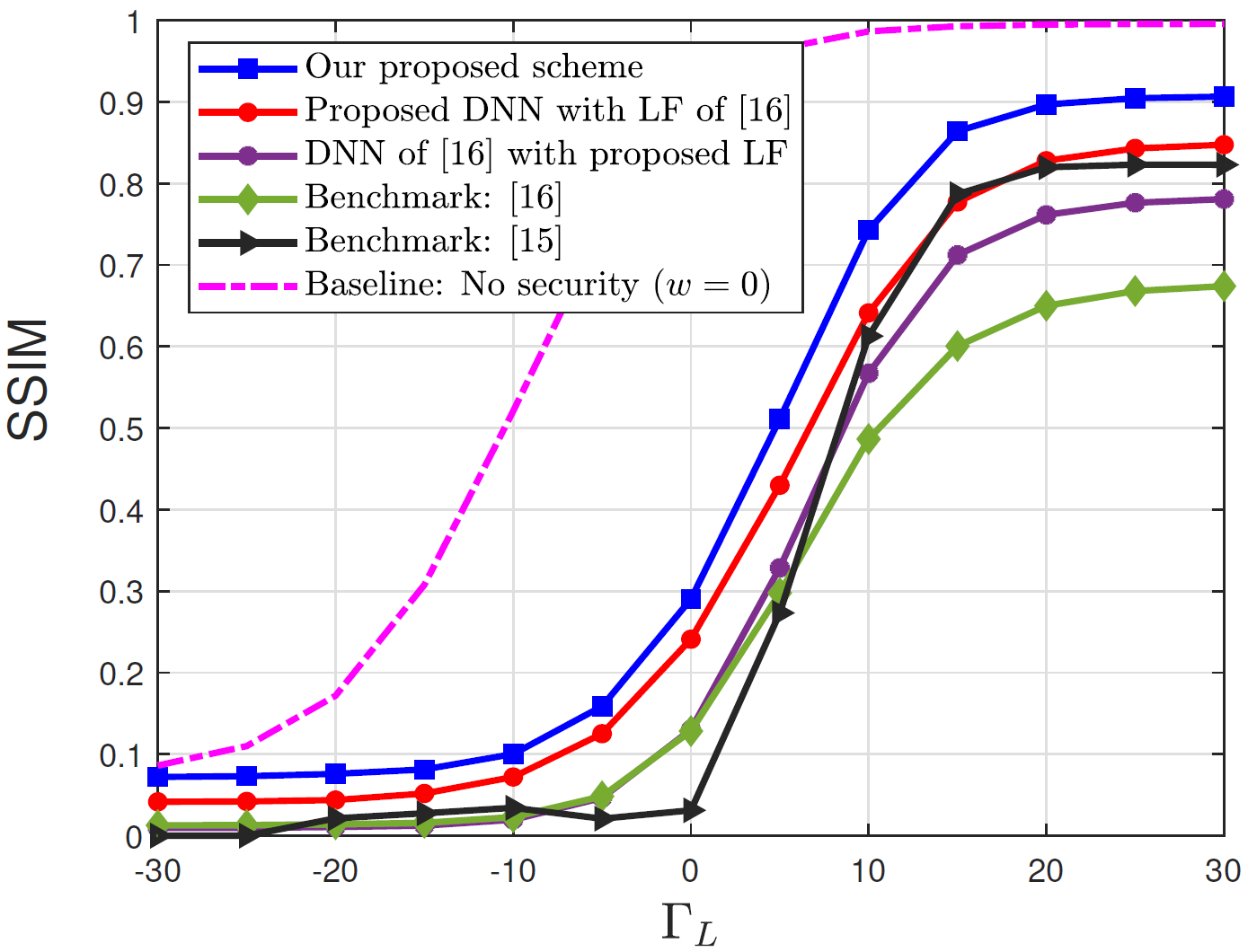}
            \vspace{-3mm}
			\caption*{(b) Reconstruction performance in terms of SSIM}
			\label{fig: ssim_ablation_test}
		\end{minipage}
		\hfill\hspace{0mm}
		\vspace{0mm}
		\caption{Performance evaluation during  inference.} 
		\label{fig:ablation_and_nT}
		\vspace{0mm}
	\end{figure*}

\begin{figure}
    \centering
			\includegraphics
			[width=3.25in,height=2.25in] 
			{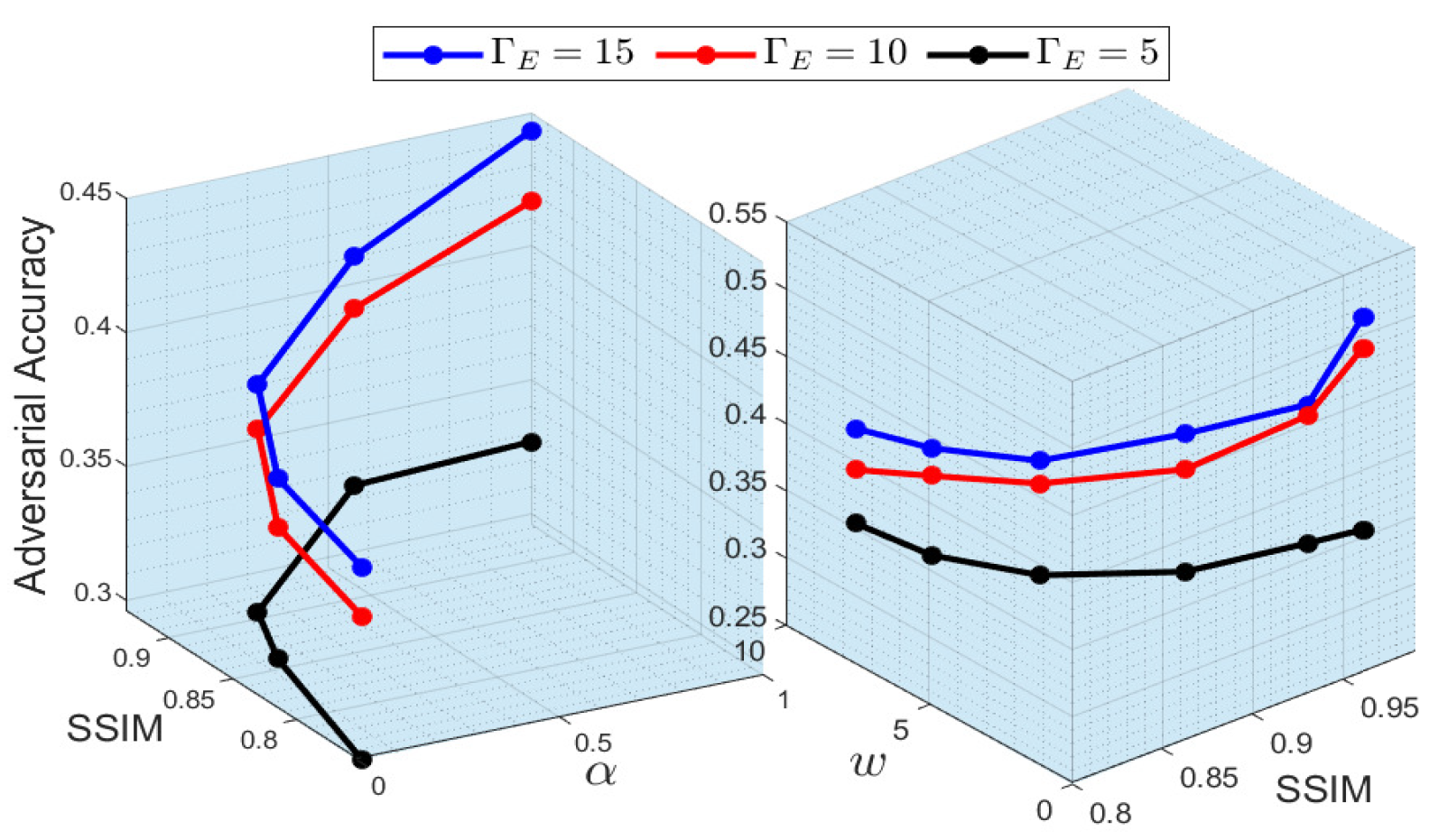}	
            \vspace{0mm}
			\caption{\textcolor{black}{Effect of $\alpha$ and $w$ on the privacy-utility trade-off}}
			\label{fig:3d_coeffs}
\end{figure}

	Fig. \ref{fig:util} demonstrates the \emph{privacy-utility trade-off} for our proposed  scheme. 
	The figure is obtained for the legitimate SNR from $\Gamma_{L}=-20$dB to $30$dB with steps of $5$ dB. 
	The figure addresses  the fact that by improving the quality of the reconstructed image in SSIM, it is inevitable to have a certain level of information leakage  in general.
	This phenomena, which  is actually in accordance with one's intuition,   also verifies our generic  mathematical framework proposed  in \eqref{eq:P0}. 
	The point of this figure is that by increasing the SNR and gradually  improving the quality of  data  reconstruction (increasing the SSIM at Bob), the information leakage soon saturates. After that, we can achieve higher SSIMs without any significant  increase in  the leakage.   For instance,  having $M=3$ adversaries, by setting $\Gamma_{L}\geq 5$ dB, we can enhance the image reconstruction at Bob's decoder, without any further decrease  in information. In this case, $15$ dB increase in channel SNR can improve the SSIM about $56\%$.  
	Therefore, such  trade-off curves can help network designers  adjust the system parameters to achieve desired levels of privacy and utility. 
	The (small) negative impact of adversaries on the  SSIM values could be explained by the minimax game invoked
    in Sections \ref{sec:training} and \ref{sec:evaluation}, which mutually affects the performance of legitimate counterparts as well.

  \subsubsection*{Adversarial strategies and eavesdropping performance evaluation:}  
 We study the performance of our proposed scheme under two different adversarial strategies: Adversaries may infer  private attributes either individually (non-colluding case), or by 
	learning from 
	the combination of the 
	individual logits
	obtained  by each  eavesdropper 
	(colluding setup).   
   Mathematically speaking,   for the non-colluding setup, the training is performed based on \eqref{eq:L_AB-LE}--\eqref{eq:distortion},  and the eavesdroppers try to extract the common $S$ or individual secret $S_i$ based on their own loss function in  \eqref{eq:L_E}.  
	Meanwhile, for the colluding setup (with a common secret $S$) an extra level of ``knowledge combination'' is performed, where adversaries employ \emph{ensemble learning} method and combine the individually-extracted logits 
    to infer the secret $S$.  That is, a weighted sum in the form of $\sum_{i\in[M]}^{} { \varphi_i q_{\Theta_{E,i}}(s_i|z_i)}$  is considered  in the training, where $\omega_i$'s are learned  by the set of colluding eavesdroppers via ensemble learning.


	Figs. 
\ref{fig:sut_and_acc}-(a)
and
\ref{fig:sut_and_acc}-(b) 
demonstrate the adversarial accuracy of our proposed scheme for the scenario of colluding eavesdroppers.
For this experiment, we study  the performance of \emph{colluding adversarial inference,} and  compare it with the non-colluding setup.
{We note that for the non-colluding setup, the training is performed based on \eqref{eq:L_AB-LE}--\eqref{eq:distortion},  and the eavesdroppers try to extract the common or individual secret $S_i$ based on their own loss functions given in  \eqref{eq:L_E}.  
	Meanwhile, for the colluding setup (with a common secret $S$) an extra level of ``knowledge combination'' is performed afterward---we combine the  individually-extracted logits of adversaries, and  
	leverage  the concept of \emph{ensemble learning} to infer the secret $S$.}  
{
	For the colluding setup, the adversarial performance is measured by the overall accuracy of adversaries in correctly finding the ground-truth $\bm \varepsilon_s$ from their aggregated logits, while for the non-colluding benchmarks, the mean accuracy across eavesdroppers is plotted for the sake of comparison.} 
The overall accuracy of adversaries 
is plotted versus channel SNR of the wiretap links,  $\Gamma_{E}$, for different number of eavesdroppers and different types of communication channels.    
	We can observe from the figures that increasing the number of eavesdroppers results in achieving  higher  accuracy for the adversaries, which is aligned with one's intuition.  
	The increase in adversarial accuracy is more significant in the colluding case, due to the collaboration and knowledge combination among eavesdroppers which  helps them learn the secret more accurately. 
	Figs. 
	\ref{fig:sut_and_acc}-(a) and \ref{fig:sut_and_acc}-(b)  
	indicate that by increasing the quality of adversarial links, i.e., increasing $\Gamma_{E}$, 
	the accuracy of adversaries  
	increases by 
	at most  $10\%$.  
	This can be observed from \eqref{eq:snr}, where  higher values for $\Gamma_{E}$ results in having less-distorted (less noisy) observations $Z_i^k = z_i$, $i \in [M]$, at the  adversarial DNNs, which results in  more accurate  estimations  about the  posterior adversarial distribution  $q_{\Theta_{E,i}}({s|z_i}),\bm$  with respect to the ground-truth  ${\bm\varepsilon}_s$.      
	Interestingly, the amount of increase in the adversarial accuracy reduces with the increase in $\Gamma_E$, which highlights the limitation of adversarial nodes based on our proposed  scheme. 
	Fig. 
	\ref{fig:sut_and_acc}-(b)  
	also addresses  the generalization capability of our learning-based  framework 
	extended to 
	the AWGN communication model and Nakagami-$m$ scenario $|h|^2 \sim \mathcal{G}(m, 1/m)$.  
	This experiment verifies that we can achieve almost similar  performance in other  channel scenarios, despite being trained for the  Rayleigh fading case, which  highlights the  
	\emph{robustness} and \emph{generalizability} of our proposed scheme.


	\vspace{-0.5mm}
	Fig.
	\ref{fig:ablation_and_nT}-(a)
	demonstrates the information leakage for our proposed  scheme   compared with different benchmarks. 
	Notably,  this  figure highlights the superiority of our approach  
	compared to the benchmarks of \cite{AE-Deniz} and \cite{Deniz-GDN} in terms of the mutual information.  
	This figure also implies that increasing the  SNR of legitimate link $\Gamma_{L}$, 
    results in having lower information leakage
	This is because increasing $\Gamma_{L}$ (given in \eqref{eq:snr}) results in  
	less noisy observations at Bob than  
	the adversarial DNNs. 
	Therefore, Alice can better hide the sensitive  attributes  in noise, making it harder for the adversaries to extract the sensitive information from  distorted received  data.   

    \vspace{-0.5mm}
	Fig. 
	\ref{fig:ablation_and_nT}-(b)
	illustrates the data   reconstruction performance of our proposed framework 
	compared with other benchmarks. 
	One can infer from this figure and Fig. 
	\ref{fig:ablation_and_nT}-(a)
	that our proposed system outperforms the benchmarks in terms of both information  leakage  and  utility. 
	Accordingly, $20\%$ and $10\%$ performance gain is achieved by our proposed scheme 
	compared with  \cite{AE-Deniz} and \cite{Deniz-GDN}, respectively.  
	Fig.
	\ref{fig:ablation_and_nT}-(b)
	also implies that increasing   $\Gamma_{L}$ results in having higher SSIM values. 
	This is because increasing $\Gamma_{L}$ results in   less distorted  observations at Bob,  which facilitates the image reconstruction performance of Bob's neural decoder.  
	Notably, Fig. 
	\ref{fig:ablation_and_nT}-(b)
	highlights that if we neglect the existence of adversaries and set $w = 0$, 
	our proposed scheme can achieve almost perfect  (${\sf SSIM} = 1$) data recovery.  
	The ablation-like examinations conducted  in  Figs. 
	\ref{fig:ablation_and_nT}-(a)
	and
	\ref{fig:ablation_and_nT}-(b) 
	imply  that both the DNN structure and the employed  LF for optimizing the 
	framework contribute to the overall  privacy  and  reconstruction  performance.  
	Specifically,  our proposed DNNs at Alice and Bob, together with the proposed objective function elaborated in  \eqref{eq:P-DL-CE}  and \eqref{eq:L_AB-LE}--\eqref{eq:distortion}  have resulted in achieving the best performance compared with other benchmarks and baselines. 
	

Fig. \ref{fig:3d_coeffs} 
   studies the impact of tuning parameters $\alpha$ and $w$  on the adversarial performance of our proposed system.  
		 These tuning parameters 
		 are
		 the coefficients associated with utility and privacy    adjustment  within our training LF in 
		 \eqref{eq:distortion} and \eqref{eq:L_AB-LE}, respectively.   
		 For this experiment, the adversarial performance is captured by investigating the accuracy of adversaries in correctly finding the ground-truth label $\bm \varepsilon_s$ (representing the  sensitive information $S$) out of $L = 10$ labels of  CIFAR-10 dataset. 
         {For the accuracy measurement in this experiment, we consider the pessimistic scenario in which the  correct guess  of true labels 
   by any adversary $i \in [M]$ 
   increments  the total accuracy of adversaries.} 
		 The figure indicates that by increasing $\alpha$, higher values of SSIM can be achieved, since more emphasis on the SSIM error $\Delta^{\sf SSIM}$ is put based on \eqref{eq:distortion}. However, the accuracy of adversaries in extracting the private  information also increases, which verifies the privacy-utility trade-off. 
		 Notably, by increasing  $\alpha$ to values  more than $0.1$, a jump in the adversarial accuracy   can be observed, which lead us choosing $\alpha = 0.1$ for our network.   
		 Similarly, by increasing $w$, the emphasis goes toward the secrecy criteria introduced  in   \eqref{eq:P0}, \eqref{eq:P-DL-CE}, and  \eqref{eq:L_AB-LE},  
		 which leads to the reduction in adversarial accuracy and achieved SSIM, verifying the privacy-utility trade-off as well.  
		The figure
		also  shows that increasing the adversary  channels SNR $\Gamma_{E}$ can improve the  adversarial accuracy in finding the sensitive data 
		 $\bm \varepsilon_s$.  Interestingly, the  amount of increase in the adversarial accuracy reduces with the increase in $\Gamma_{E}$ which highlights the limitation of adversarial nodes based on our proposed  scheme.

    \begin{figure}
        \centering
        \includegraphics
			[width=3.5in,height=2.5in, 
			trim={0in 0.0in 0 0.0in},clip] 
			{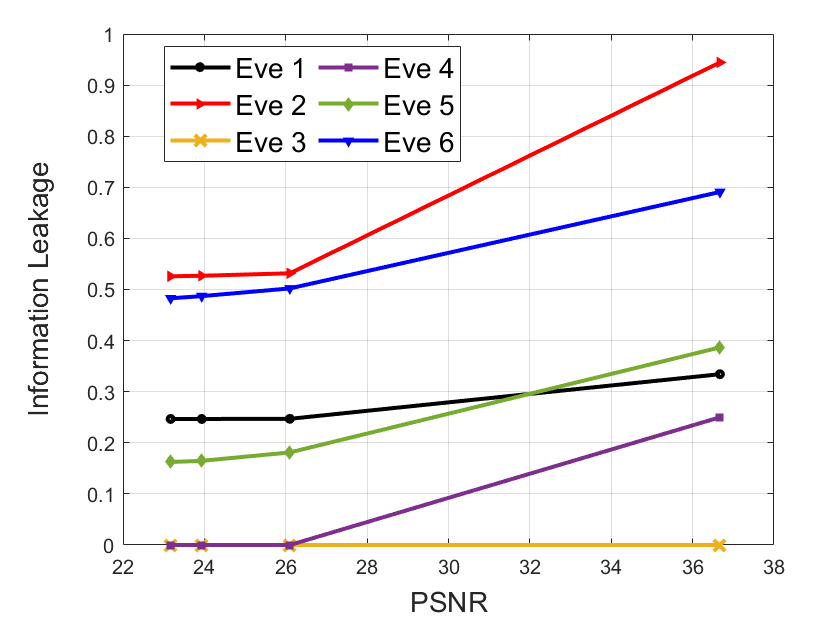}
			\caption{\textcolor{black}{Privacy-utility trade-off over CelebA dataset.}}
			\label{fig:celebA_PUT}
    \end{figure}

\begin{figure}
    \centering
   \includegraphics
			[width=3.5in,height=2.5in, 
			trim={0.0in 0.0in 0 0.0in},clip] 
			{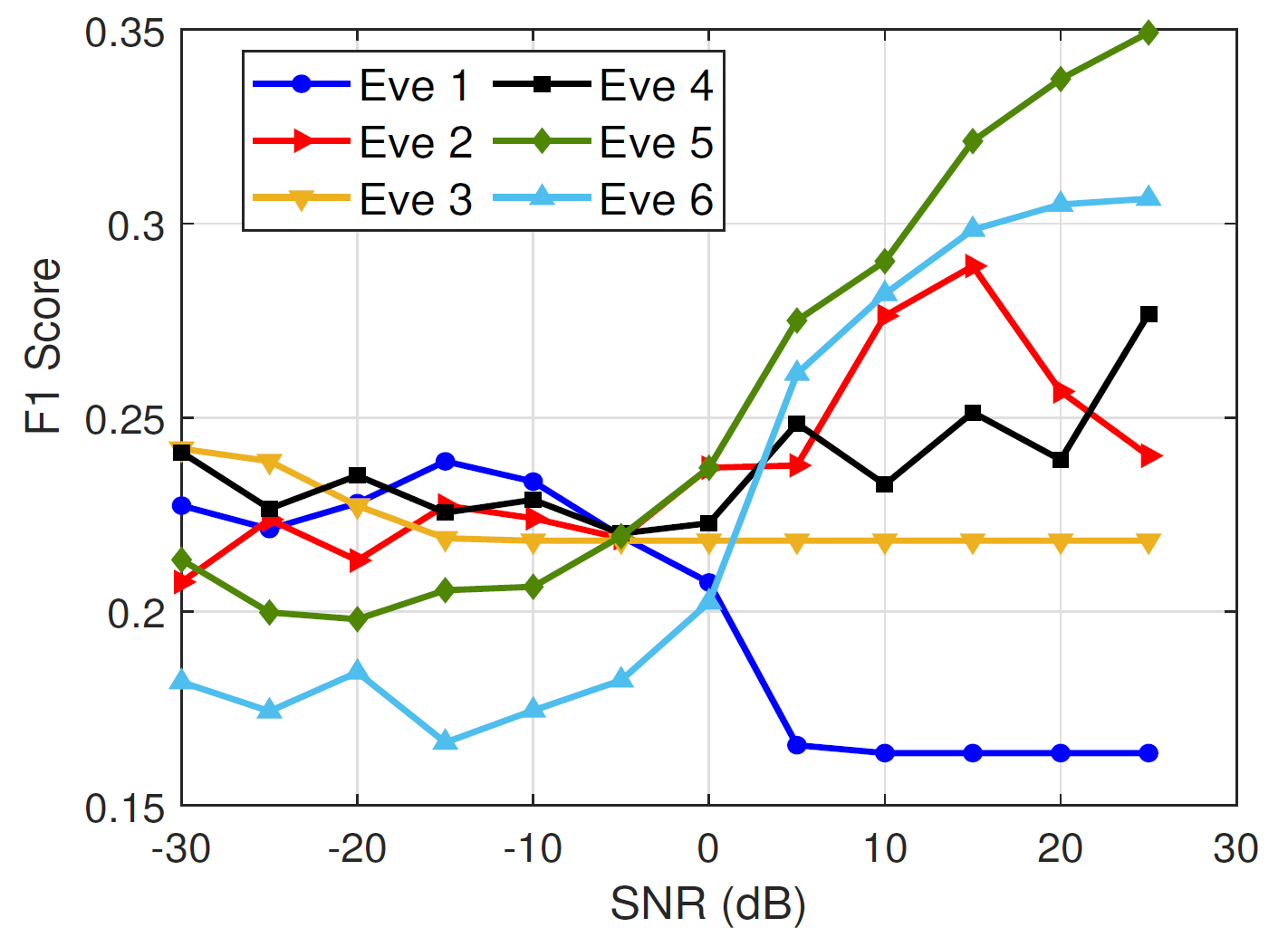}
			\caption{Adversarial accuracy in terms of F1 score over CelebA dataset.}
			\label{fig:celebA_adv_acc}
\end{figure}

	\vspace{0mm}
	\begin{table}[tbph!]
		\centering
		\small
		\caption{\small {Privatized CelebA Attributes for Each Eavesdropper}}\label{appendix-celebA}
		\vspace{0mm}
		\begin{tabular} 
			{|p{0.7in}|p{2.5in}|}
			\hline \textbf{Eavesdropper} & \textbf{Private Attributes}\\
			\hline
			\hline
			Eve $1$ & {\fontfamily{qcr}\selectfont
				Wavy\underline{\hspace{2mm}}Hair}
			and 
			{\fontfamily{qcr}\selectfont Black\underline{\hspace{2mm}}Hair}
			\\
			Eve $2$ & {\fontfamily{qcr}\selectfont
				Wearing\underline{\hspace{2mm}}Lipstick}
			and 
			{\fontfamily{qcr}\selectfont Smiling}\\
			Eve $3$ & {\fontfamily{qcr}\selectfont
				Double\underline{\hspace{2mm}}Chin}
			and 
			{\fontfamily{qcr}\selectfont Wearing\underline{\hspace{2mm}}Necklace}\\
			Eve $4$ & {\fontfamily{qcr}\selectfont
				No\underline{\hspace{2mm}}Beard}
			and 
			{\fontfamily{qcr}\selectfont 5\underline{\hspace{2mm}}o\underline{\hspace{2mm}}Clock\underline{\hspace{2mm}}Shadow}\\
			Eve $5$ & {\fontfamily{qcr}\selectfont
				Bags\underline{\hspace{2mm}}Under\underline{\hspace{2mm}}Eyes}
			and 
			{\fontfamily{qcr}\selectfont Arched\underline{\hspace{2mm}}Eyebrows}\\
			Eve $6$ & {\fontfamily{qcr}\selectfont High\underline{\hspace{2mm}}Cheekbones}
			and 
			{\fontfamily{qcr}\selectfont Pointy\underline{\hspace{2mm}}Nose}\\
			\hline
		\end{tabular} \vspace{-5mm}	
	\end{table}	
	
	\vspace{-1mm}
	\subsection{Evaluation on CelebA dataset}	 
	In this subsection, we evaluate our proposed scheme  on CelebA  dataset \cite{CelebA}. 
	This is a large-scale face attributes dataset with more than 200k images of celebrities with 40 attribute annotations.  
	For the following experiments, we consider the non-colluding scenario, in which eavesdroppers are interested in  \emph{different  secrets} $S_i \in \mathcal{S}_i$, while we  assume $|S_i| = 4$ for all eavesdroppers.   Details of the 
	privatized attributes for each eavesdropper $i\in [M]$ 
	are provided in Table  \ref{appendix-celebA}.

	Fig. \ref{fig:celebA_PUT} demonstrates the privacy-utility trade-off over CelebA dataset with SNR $\Gamma_{E} =25$dB. Since the eavesdroppers are interested in different secrets,  the results are plotted 
	for each individual eavesdropper $i \in [6]$.   
	The figure  is obtained by changing the hyperparameter $w \in \{0, 10, 50, 100\}$. 
	In this figure, the extreme case of $w=100$ 
	has the minimum leakage
    while having the minimum utility (measured by the PSNR in this experiment). 
	This is because 
	larger $w$ implies that DNN encoder-decoder pair gives more importance to  the secrecy criteria 
	in \eqref{eq:P-DL-CE}--\eqref{eq:L_AB-LE}.  
	However, as we gradually decrease $w$,    higher PSNRs can be achieved with a slight increase in the information leakage.   
	The figure also implies that for the baseline of no security design, i.e., $w=0$, the information leakage increases by about $30 \%$, which  highlights the importance of designing  a secure neural encoding-decoding pair.    

\textcolor{black}{
\emph{Remark 5:} 
We remark that the privacy-utility trade-off curves enable users to dynamically select the \emph{operation point} that matched their specific requirements, balancing between utility and privacy. We
quantified this balance with hyperparameter $w$,  where our experiments demonstrate how different $w$ values affect this balance - higher values favor utility (image quality) while lower values prioritize privacy. 
}

	{Fig. \ref{fig:celebA_adv_acc} shows the accuracy of eavesdroppers versus $\Gamma_{E}$ for $w=10$.  
		Since  CelebA  is a highly \emph{imbalanced dataset} with respect to most of the attributes, we have chosen the F-score metric with macro averaging to measure the  adversarial accuracy  in this scenario.
      Note that   due to the imbalance in CelebA dataset attributes, we employed a penalty mechanism. However, we still observe the effect of the imbalance on the eavesdroppers in terms of F-scores.
		The figure  indicates that our privacy-aware  neural encoding-decoding pair has been successful in confusing Eves about the private attributes, as a maximum of 25\% accuracy is achieved by adversaries for most of the SNR values. Similar to the results over CIFAR-10, this figure also indicates that  increasing the quality of adversarial links can increase    the accuracy of eavesdroppers by at most  $15\%$.  
	}

\begin{figure}
    \centering
  \includegraphics
			[width=3.5in,height=2.8in]
			{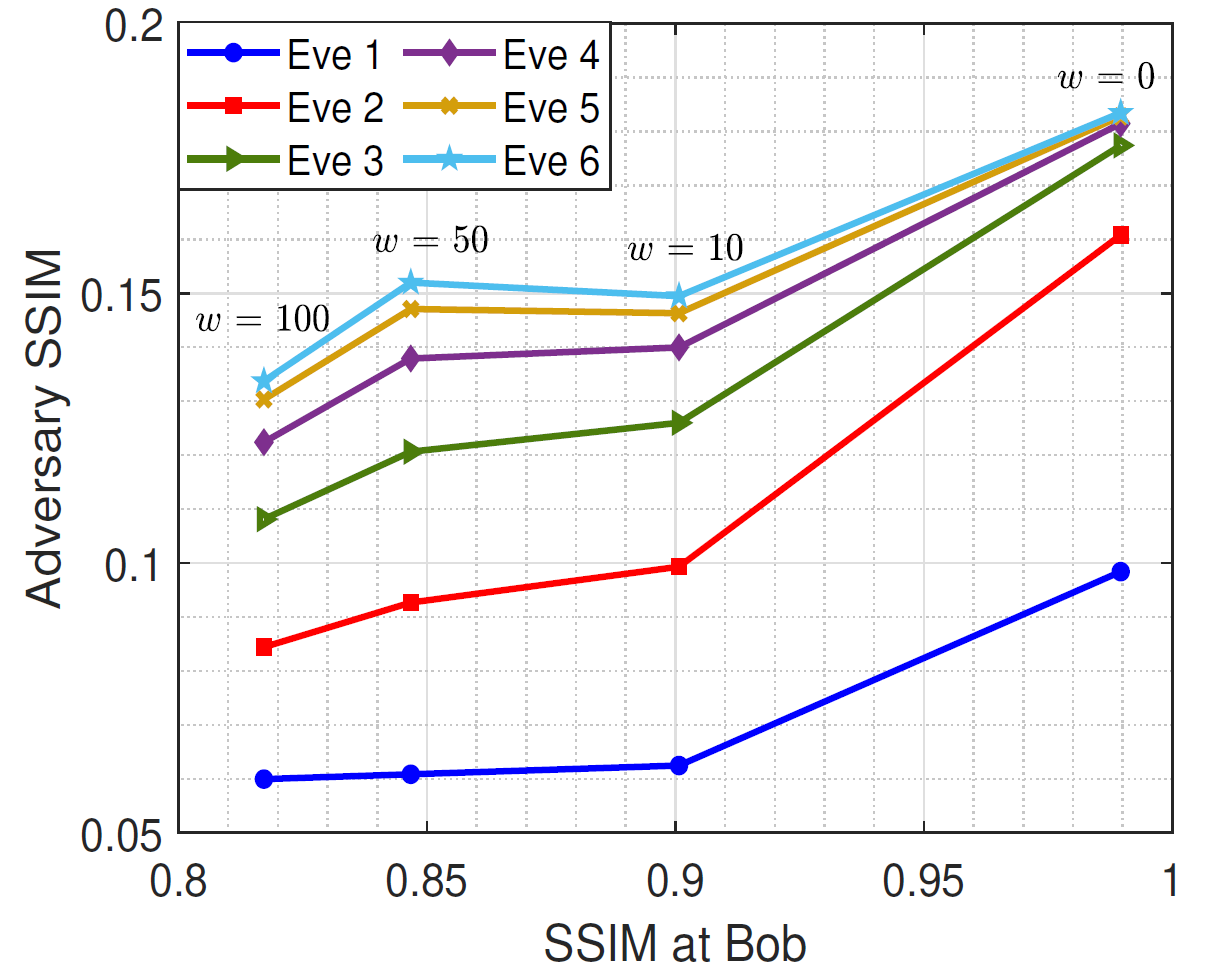}
			\caption{Reconstruction quality of the legitimate destination vs. the eavesdroppers over CelebA dataset.}
			\label{fig:celebA_adv_leg_ssim}	
\end{figure}

	{Fig. \ref{fig:celebA_adv_leg_ssim}   illustrates the reconstruction quality (for different values of $w$) at  Bob and each of the eavesdroppers. 
		We can observe that the bigger $w$, the worse would be the reconstruction quality at Eves (and also at Bob), as the emphasis shifts toward the secrecy criteria.  
		The figure also implies that the eavesdroppers are strongly restrained from reconstructing any meaningful (high perceptual quality) images, due to very low adversarial  SSIM values.   
        }

	\vspace{-0mm}
	\begin{figure}
		\centering		
        \renewcommand{\arraystretch}{0.9}
		\begin{tabular}{c} 
			\multicolumn{1}{c}{\small\textbf{Original Images}} \\
			\includegraphics[width=0.43\textwidth]{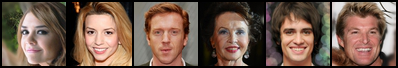}   
	       \\
			\multicolumn{1}{c}{\small\textbf{Reconstructions at Bob: 24.59dB/0.89, 22.19dB/0.81}} \\ 
            \renewcommand{\arraystretch}{0}
			\begin{tabular}{c}
				\includegraphics[width=0.43\textwidth]{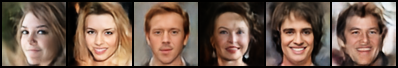}
				\\
				\includegraphics[width=0.43\textwidth]{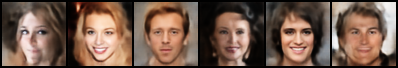}
			\end{tabular} 
			\\
			\multicolumn{1}{c}{\small\textbf{Reconstructions at Eve 1: 13.51dB/0.34, 13.75dB/0.36}} \\ 
            \renewcommand{\arraystretch}{0}
			\begin{tabular}{c}
				\includegraphics[width=0.43\textwidth]{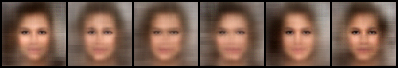}\\
				\includegraphics[width=0.43\textwidth]{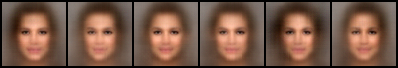}
			\end{tabular}
			\\
			\multicolumn{1}{c}{\small\textbf{Reconstructions at Eve 2: 16.81dB/0.50, 14.39dB/0.41}} \\ 
            \renewcommand{\arraystretch}{0}
			\begin{tabular}{c}
				\includegraphics[width=0.43\textwidth]{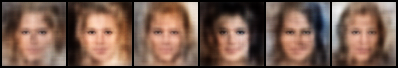}\\
				\includegraphics[width=0.43\textwidth]{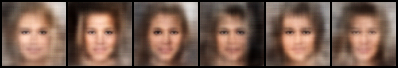} \\
			\end{tabular}
			\\
			\multicolumn{1}{c}{\small\textbf{Reconstructions at Eve 3: 19.39dB/0.66, 18.54dB/0.58}} \\
            \renewcommand{\arraystretch}{0}
			\begin{tabular}{c}
				\includegraphics[width=0.43\textwidth]{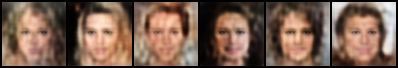}\\
				\includegraphics[width=0.43\textwidth]{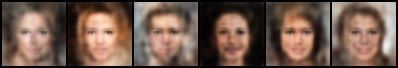} \\
			\end{tabular}
			\\
			\multicolumn{1}{c}{\small\textbf{Reconstructions at Eve 4: 19.08dB/0.66, 17.58dB/0.58}} \\
            \renewcommand{\arraystretch}{0}
			\begin{tabular}{c}
				\includegraphics[width=0.43\textwidth]{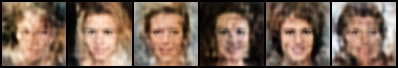}\\
				\includegraphics[width=0.43\textwidth]{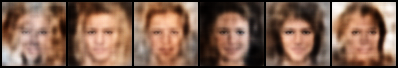} \\
			\end{tabular}
			\\
			\multicolumn{1}{c}{\small\textbf{Reconstructions at Eve 5: 21.25dB/0.74, 19.80dB/0.65}} \\
            \renewcommand{\arraystretch}{0}
			\begin{tabular}{c}
				\includegraphics[width=0.43\textwidth]{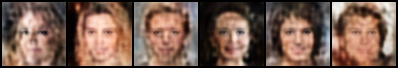}\\
				\includegraphics[width=0.43\textwidth]{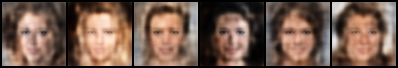} \\ 
			\end{tabular} 
			\\ 
			\multicolumn{1}{c}{\small\textbf{Reconstructions at Eve 6: 22.41dB/0.78, 21.12dB/0.71}} \\	
            \renewcommand{\arraystretch}{0}
			\begin{tabular}{c}
				\includegraphics[width=0.43\textwidth]{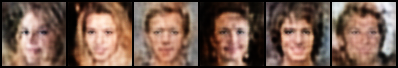}\\
				\includegraphics[width=0.43\textwidth]{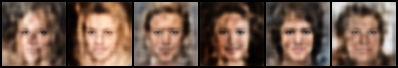} \\
			\end{tabular} 
		\end{tabular}
		\vspace{-1mm}
		\caption{Reconstructed CelebA images at Bob and each Eve.
			For each set of images, the first and second rows correspond to $w$=$10$ and $w$=$100$, respectively.  
			PSNR(dB)/SSIM values are provided above each set of images.}
		\label{fig:visual_CelebA}
		\vspace{-3mm}
	\end{figure}

	{To gain visual insights into the  reconstruction performance of Bob, and what  the reconstructed images at eavesdroppers would be like, 
		Fig. \ref{fig:visual_CelebA} is provided. 
		Please note that the goal of eavesdroppers is to infer the secrets rather than reconstruct the entire image. Here,  we show the reconstructions at the eavesdroppers only for illustration purposes.      
		To this end, the legitimate DNN encoder is frozen after being trained for $180$ epochs, and then, a similar neural structure to that of Bob is considered for each of the eavesdroppers, which gets trained for image reconstruction.\footnote{During the evaluation,  we transmit each image 10 times  and the performance metrics are  averaged over these  realizations.
		} 
		The reconstructed images in Fig. \ref{fig:visual_CelebA} correspond to the default case of $w=10$, and the extreme case of $w=100$.  
		Interestingly, one can observe that our proposed  framework has 
		been successful in hiding 
		the corresponding private information  against each eavesdropper, e.g., the hair style, smiling, facial attributes, etc.,  while recovering the entire image at Bob such that it can be distinguished clearly. 
		Notably, although the main objective of the legitimate encoder was to prevent adversaries from extracting the privatized attributes, we   observe that the eavesdroppers cannot successfully reconstruct the  images either.}


	\vspace{0mm}
	\section{Conclusions {\& Future Directions}}\label{sec:Conclusion}
	\vspace{0mm}
	In this paper, we proposed  a privacy-aware  DeepJSCC framework  for end-to-end  image transmission  in the presence of multiple eavesdroppers.    
    DeepJSCC pipeline  was designed to transmit images over the communication channel with the highest reconstruction fidelity, while   controlling  the privacy leakage theoretically 
	measured by the mutual information metric. 
Autoencoders (with convolutional neural networks as backbone)  were  implemented for neural encoder-decoder pair, along  with  an information-theoretic ``privacy  funnel'' framework  as the training loss function, realizing the privacy-utility trade-off while taking perceptual  quality  into account.   
    Extensive numerical experiments were carried out over CIFAR-10 and CelebA dataset. 
	 	Ablation studies and data visualizations have been proposed to  validate the performance gain of our  scheme compared with related benchmarks,  both numerically and visually. 
\textcolor{black}{While datasets like CIFAR and CelebA seem to be useful for testing  specific downstream tasks like classification or retrieval, in order to ensure out-of-distribution (OOD) generalization,  the system might need to be trained 
on richer datasets  suited specifically for image compression and decompression tasks.  Studying  OOD performance under general-purpose datasets 
can  be studied in the future works.    
   }

  Some of the key future directions in the context of DeepJSCC and information-theoretic privacy can be outlined below.   
    \begin{enumerate}
        \item \textit{Training over real-world measurement data:} 
In this paper,  we considered  known channel models to generate samples from conditional channel distributions. One could
easily drop this assumption if having enough channel state information (CSI) measurements collected from a particular wireless channel with unknown statistics. 
However, training wireless 
systems over real-world  wireless channels is the relatively long coherence time
compared to the rate at which data samples can be processed for training. Thus, a few channel realizations are observed over every mini-batch which may lead to
poor sample efficiency and slow convergence. This could be addressed in future works. 
{
\item \textit{Robustness against advanced adversarial strategies:}
In this paper we used adversarial likelihood compensation to push
adversaries toward uniform predictions. Another direction to study would be a scenario in which  adversaries employ more
sophisticated strategies like reinforcement learning or meta-learning. Such scenarios  reflect the ``asymmetric information'' case from game-theoretic perspective, while in this paper  we considered the ``complete information'' case, where both sides know each
other’s strategy. Our approach leads to Nash Equilibrium, and   both parties can optimize their
best responses resulting in more predictable and stable outcomes. However, asymmetric
information scenario may lead to more dynamic and adaptive game playing and creates an uncertainty which can be studied in the future works.
}
\item \textit{The role of novel generative models:} Integrating novel generative models, such as denoising diffusion probabilistic models, into DeepJSCC frameworks can help enhance  the robustness of wireless  system against adversarial attacks.  Diffusion models, with their underlying capability to learn complicated distributions and generate content out of noisy signals, have been shown to be effective in  multimedia transmission \cite{CDiff},  removing unwanted signals in multi-user systems \cite{cFree}, while taking practical assumptions of wireless systems into account \cite{hwi}.  An interesting direction would be to study the potential roles of such generative models at the encoder side, and also the decoder side, investigating whether such generative foundation models can substitute encoder-dependent  decoders,  and realize a more generalized  model-agnostic multimedia transmission framework.    
    \end{enumerate}

\begin{figure*}
		\centering
		\begin{minipage}{0.33\textwidth}
			\includegraphics
			[width=2.25in,height=1.8in,
			trim={0.0in 0.0in 0 0.0in},clip]
			{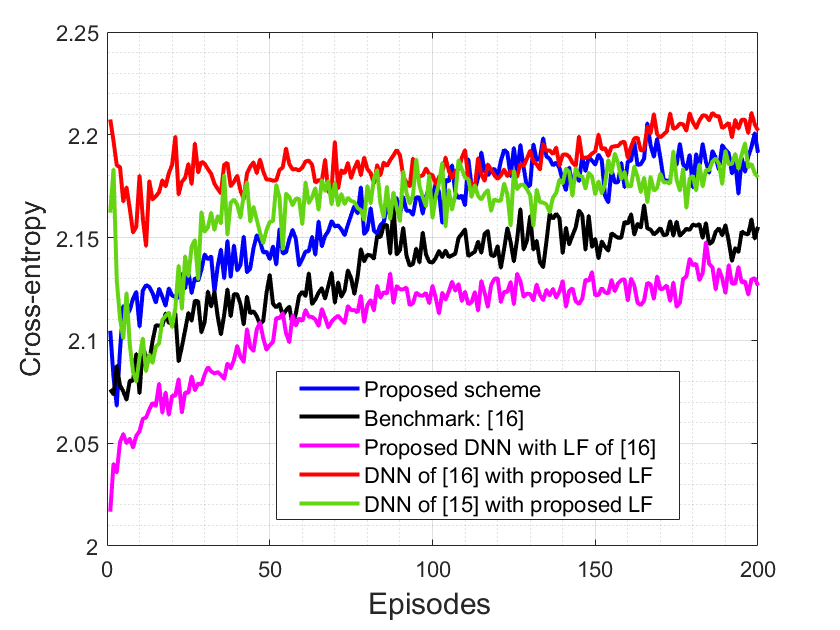}
			\vspace{0mm}
		\end{minipage}\hfill
		\begin{minipage}{0.33\textwidth}
			\includegraphics
			[width=2.25in,height=1.8in,
			trim={0.0in 0.0in 0 0.0in},clip]
			{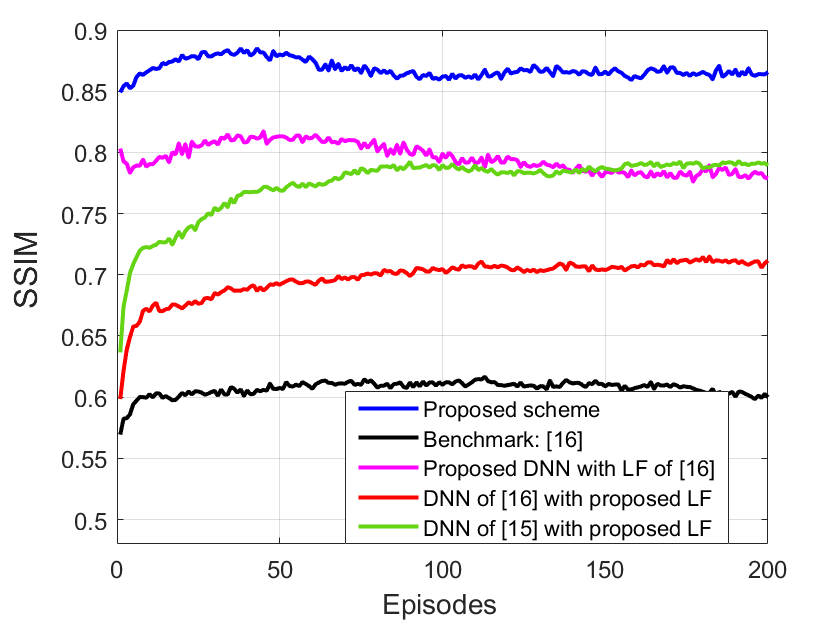}
			\vspace{0mm}
		\end{minipage}\hfill
		\begin{minipage}{0.33\textwidth}
			\includegraphics
			[width=2.25in,height=1.8in, 
			trim={0.0in 0.0in 0 0.0in},clip]
			{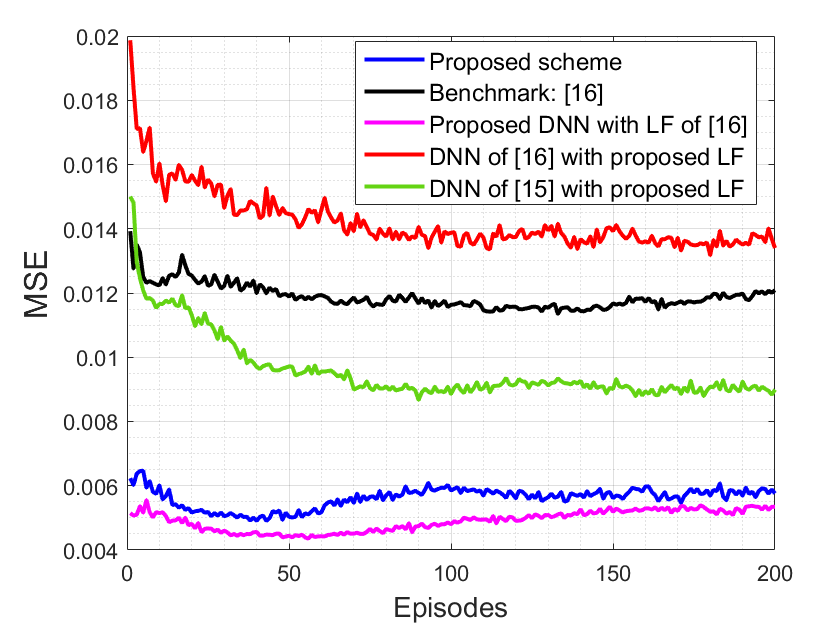}
			\vspace{0mm}
		\end{minipage}
        \vspace{0mm}
	\caption{The training process and ablation studies. 
  The cross-entropy, SSIM, and MSE metrics are visualized  from left to right.}	
		\label{fig: TP} 
  \vspace{-3mm}
	\end{figure*}

\begin{figure*}
\hspace{-5mm}		
\begin{tabular}{cccccc}
\textbf{\small Proposed Scheme} & 
\textbf{\small Benchmark 1} & 
\textbf{\small Benchmark 2} & 
\textbf{\small Benchmark 3} &
\textbf{\small Benchmark 4} &
\textbf{\small Benchmark 5}\\
			
\includegraphics[width=0.15\textwidth]{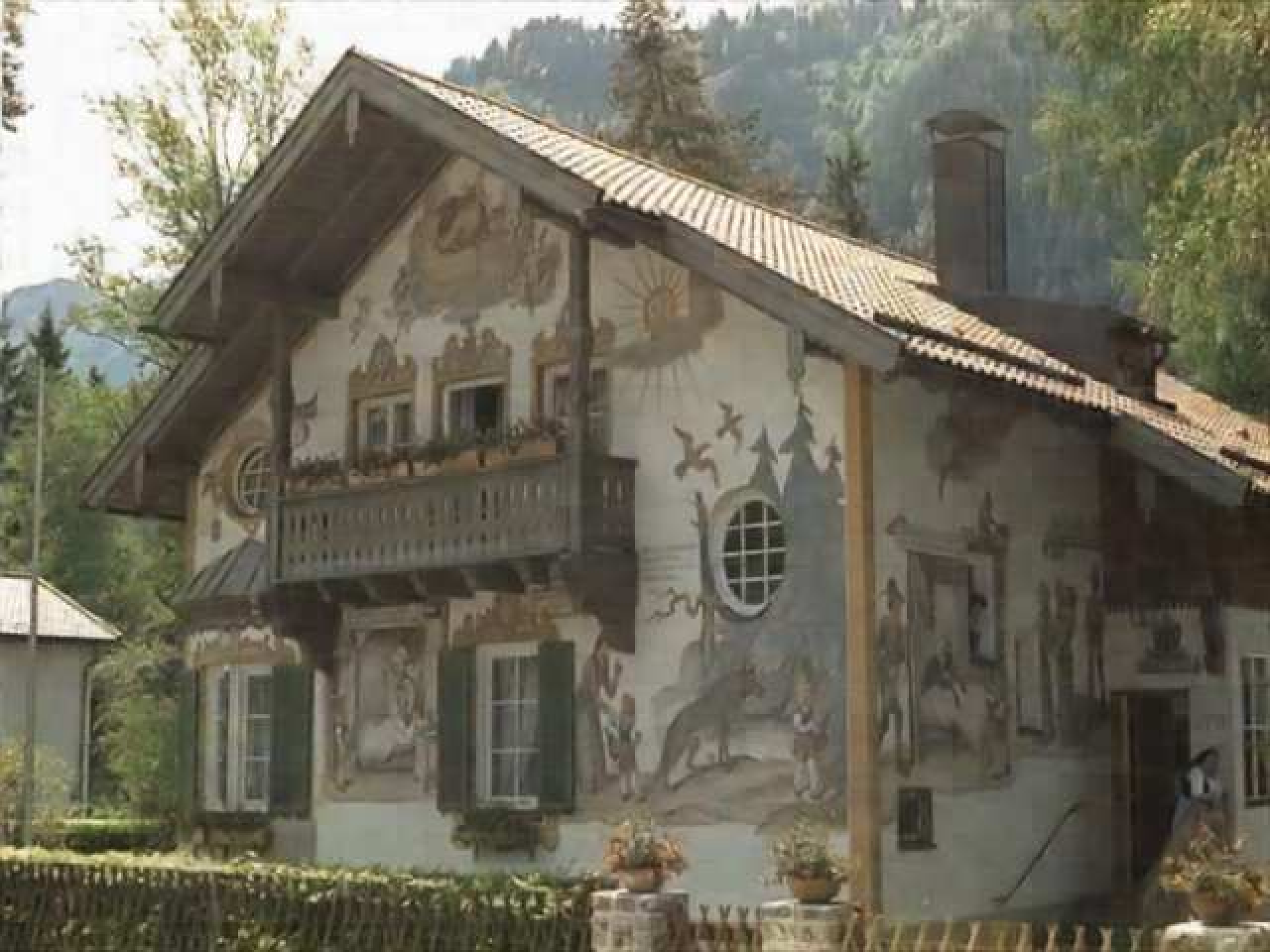}  &
\includegraphics[width=0.15\textwidth]{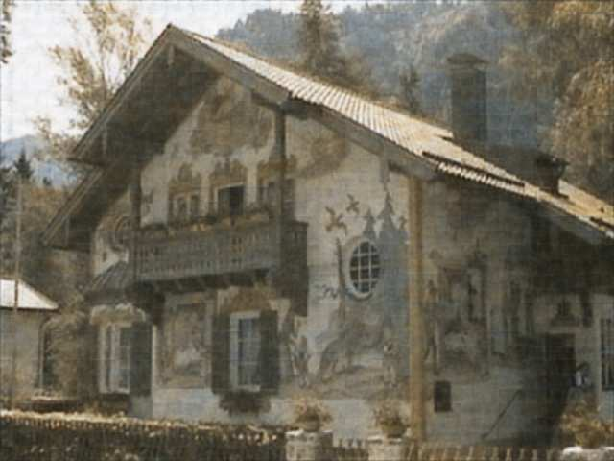} &
\includegraphics[width=0.15\textwidth]{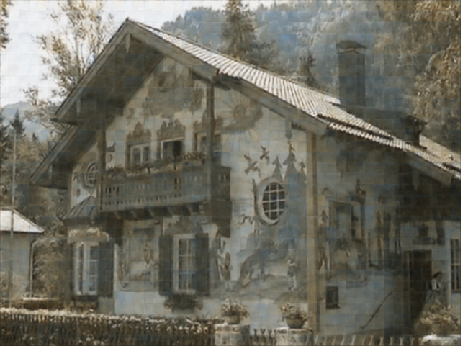} &
\includegraphics[width=0.15\textwidth]{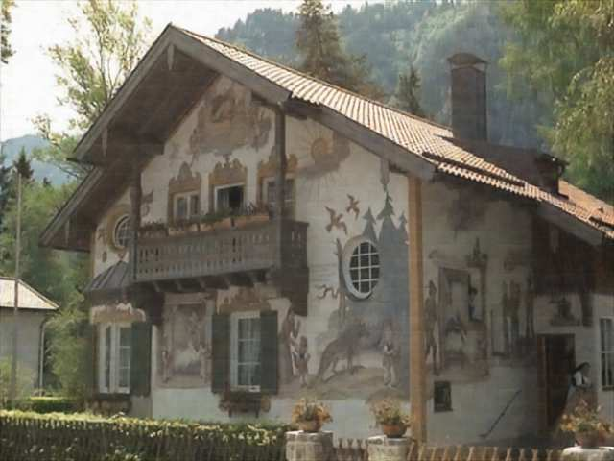}&
\includegraphics[width=0.15\textwidth]{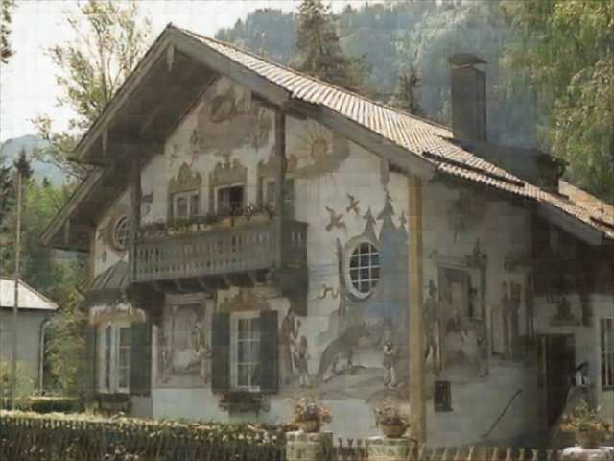}&
\includegraphics[width=0.15\textwidth]{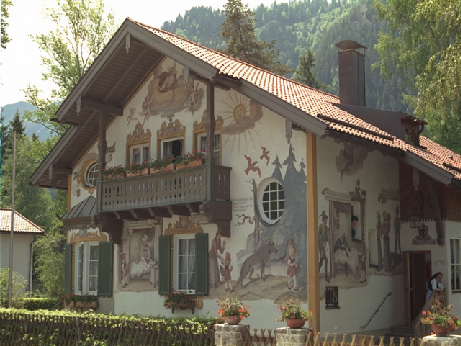}
\\
\small{26.41dB/0.85} & 
\small{22.81dB/0.64} &
\small{23.03dB/0.71} &
\small{25.73dB/0.78} &
\small{24.35dB/0.77} &
\small{39.86dB/0.99} 

\\

\includegraphics[width=0.15\textwidth]{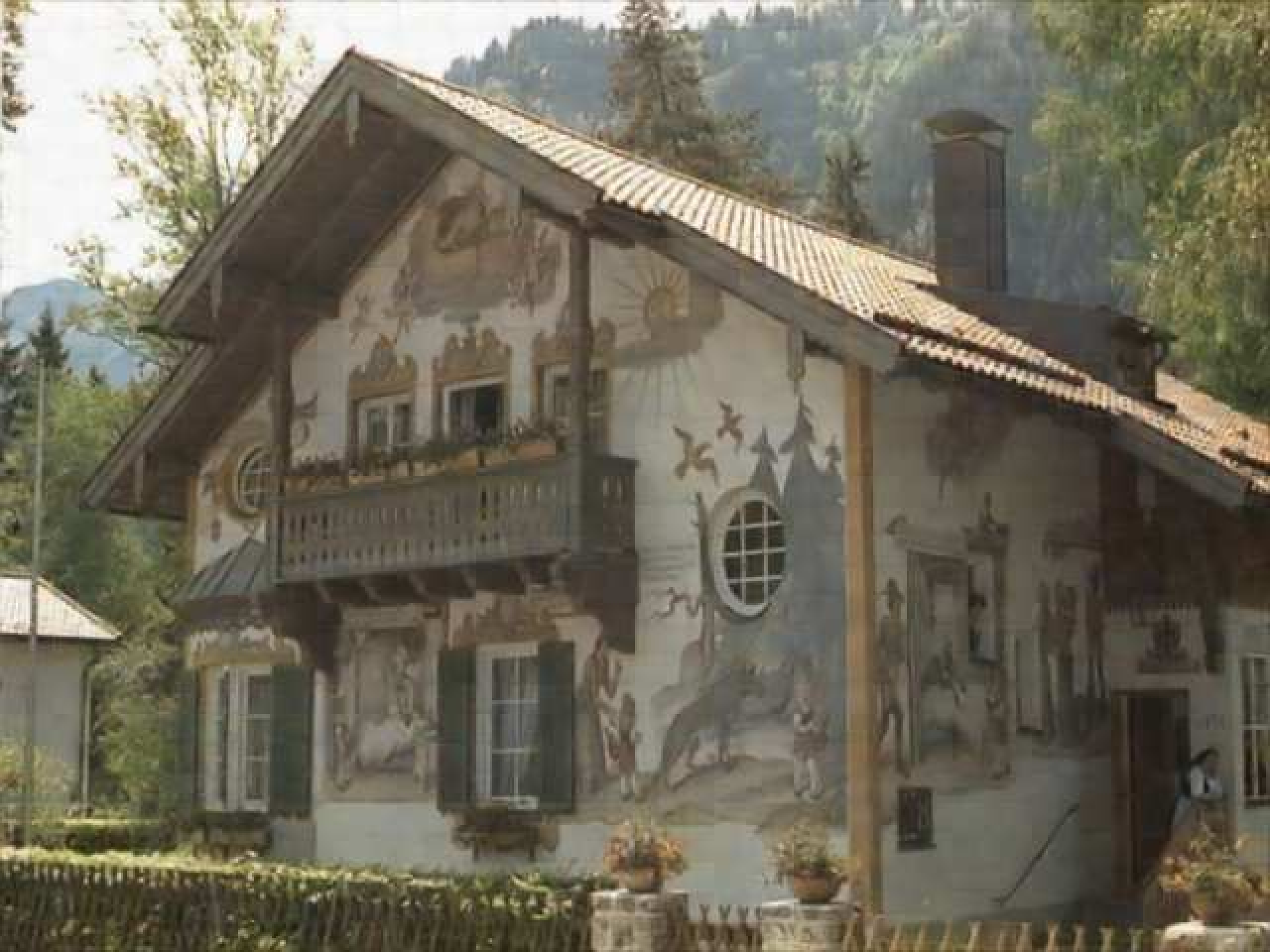}  &
\includegraphics[width=0.15\textwidth]{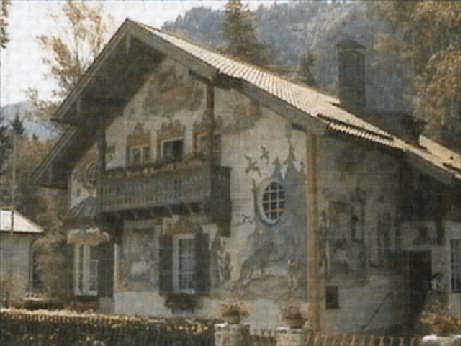} &
\includegraphics[width=0.15\textwidth]{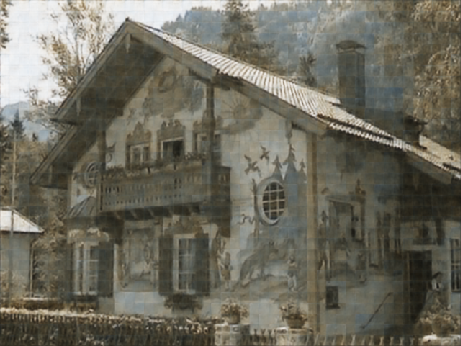} &
\includegraphics[width=0.15\textwidth]{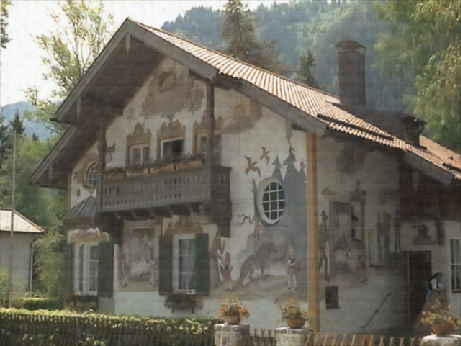}&
\includegraphics[width=0.15\textwidth]{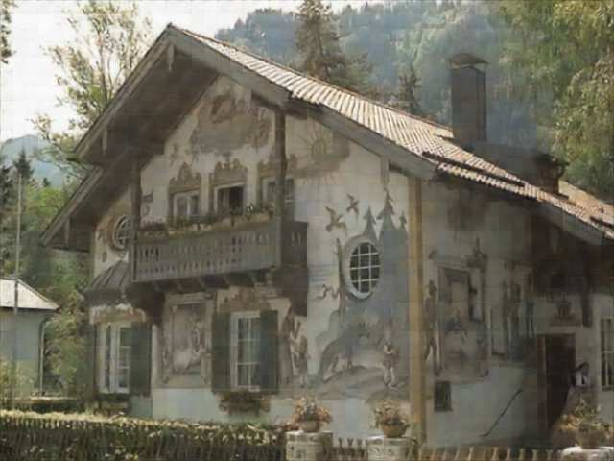}&
\includegraphics[width=0.15\textwidth]{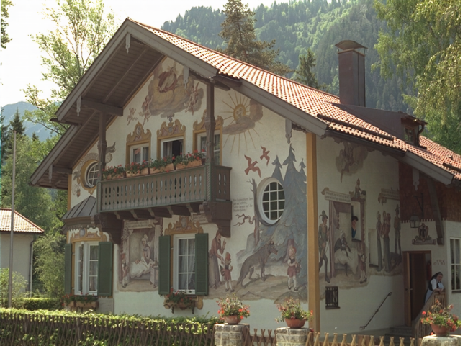}
\\
\small{25.97dB/0.84} &
\small{22.75dB/0.64} & \small{23.06dB/0.71} & 
\small{25.13dB/0.76} &
\small{24.37dB/0.77} &
\small{38.49dB/0.99}

\\

\includegraphics[width=0.15\textwidth]{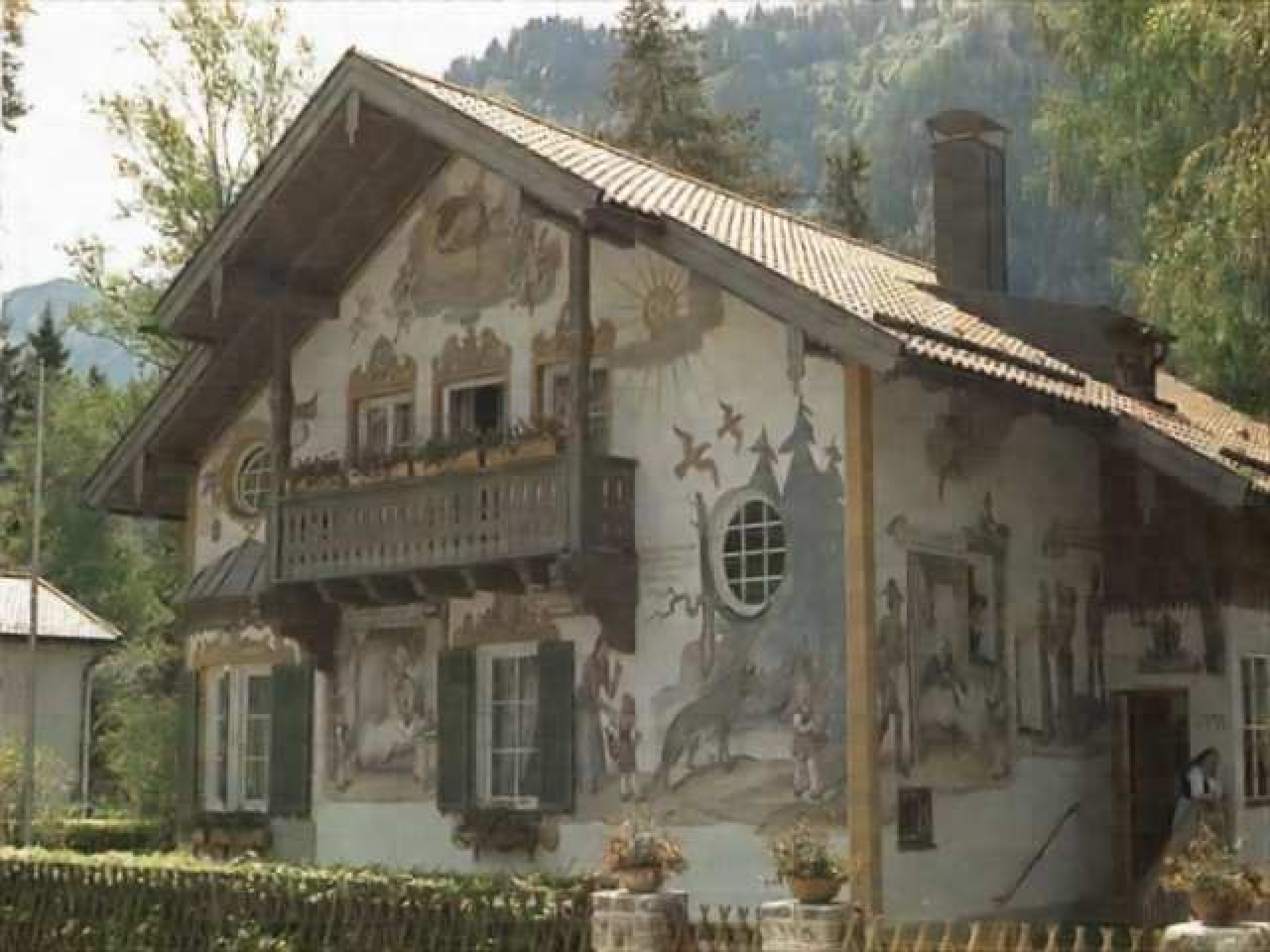}  &
\includegraphics[width=0.15\textwidth]{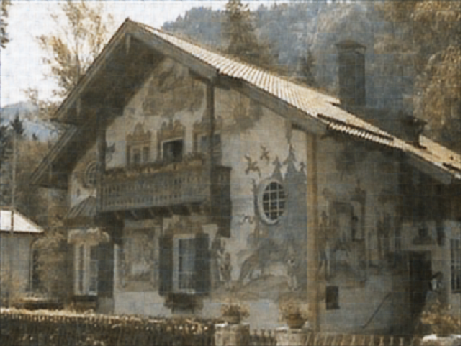} &
\includegraphics[width=0.15\textwidth]{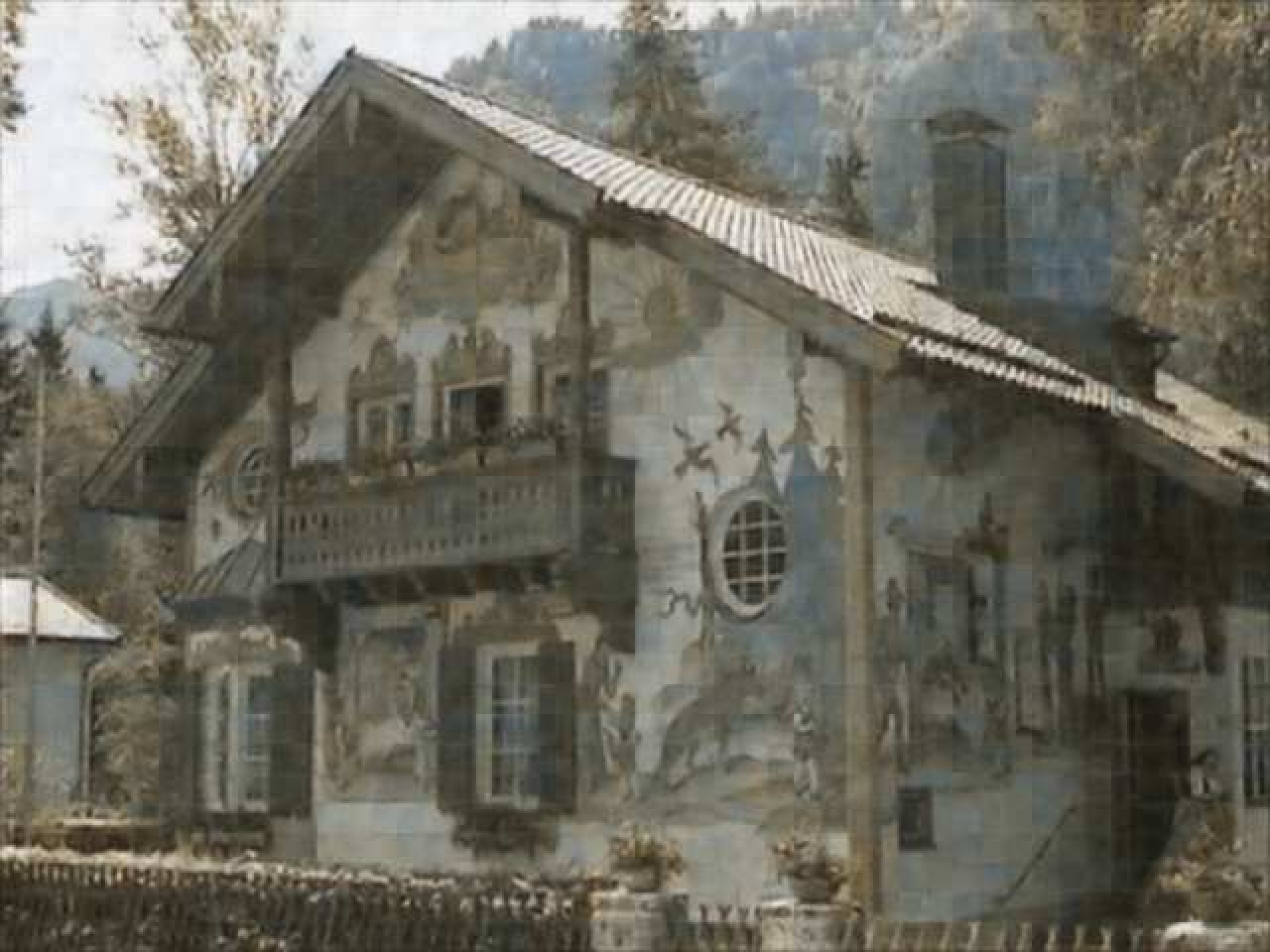} &
\includegraphics[width=0.15\textwidth]{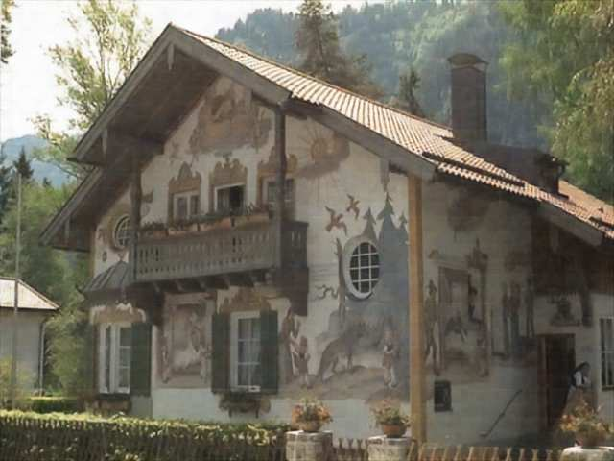}&
\includegraphics[width=0.15\textwidth]{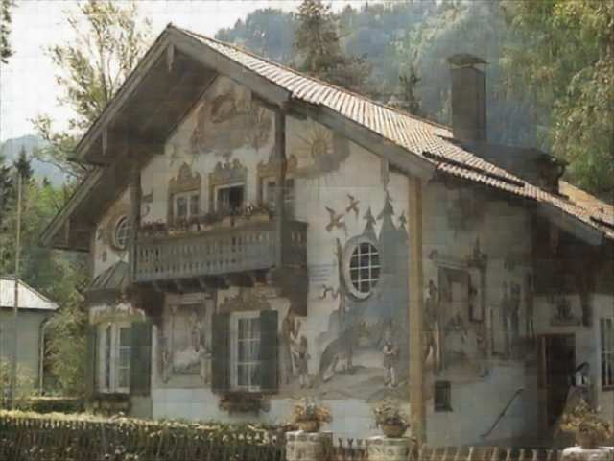}&
\includegraphics[width=0.15\textwidth]{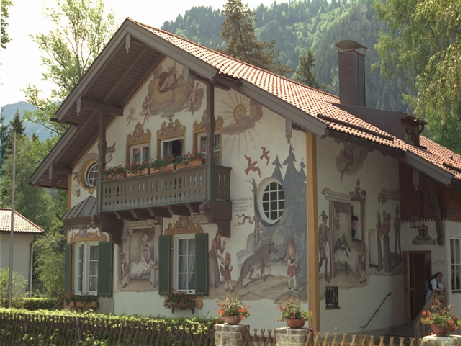}
\\
\small{26.39dB/0.85} &
\small{22.80dB/0.64} & 
\small{23.06dB/0.71} & 
\small{25.74dB/0.78} &
\small{24.32dB/0.77} &
\small{39.90dB/0.99}
			
		\end{tabular}
\vspace{0mm}
	\label{fig:visual_kodak}
\end{figure*}

\begin{figure*}
	\hspace{-5mm}		
	\begin{tabular}{cccccc}
		\textbf{\small Proposed Scheme} & 
		\textbf{\small Benchmark 1} & 
		\textbf{\small Benchmark 2} & 
		\textbf{\small Benchmark 3} &
		\textbf{\small Benchmark 4} &
		\textbf{\small Benchmark 5}\\
		
		\includegraphics[width=0.15\textwidth]{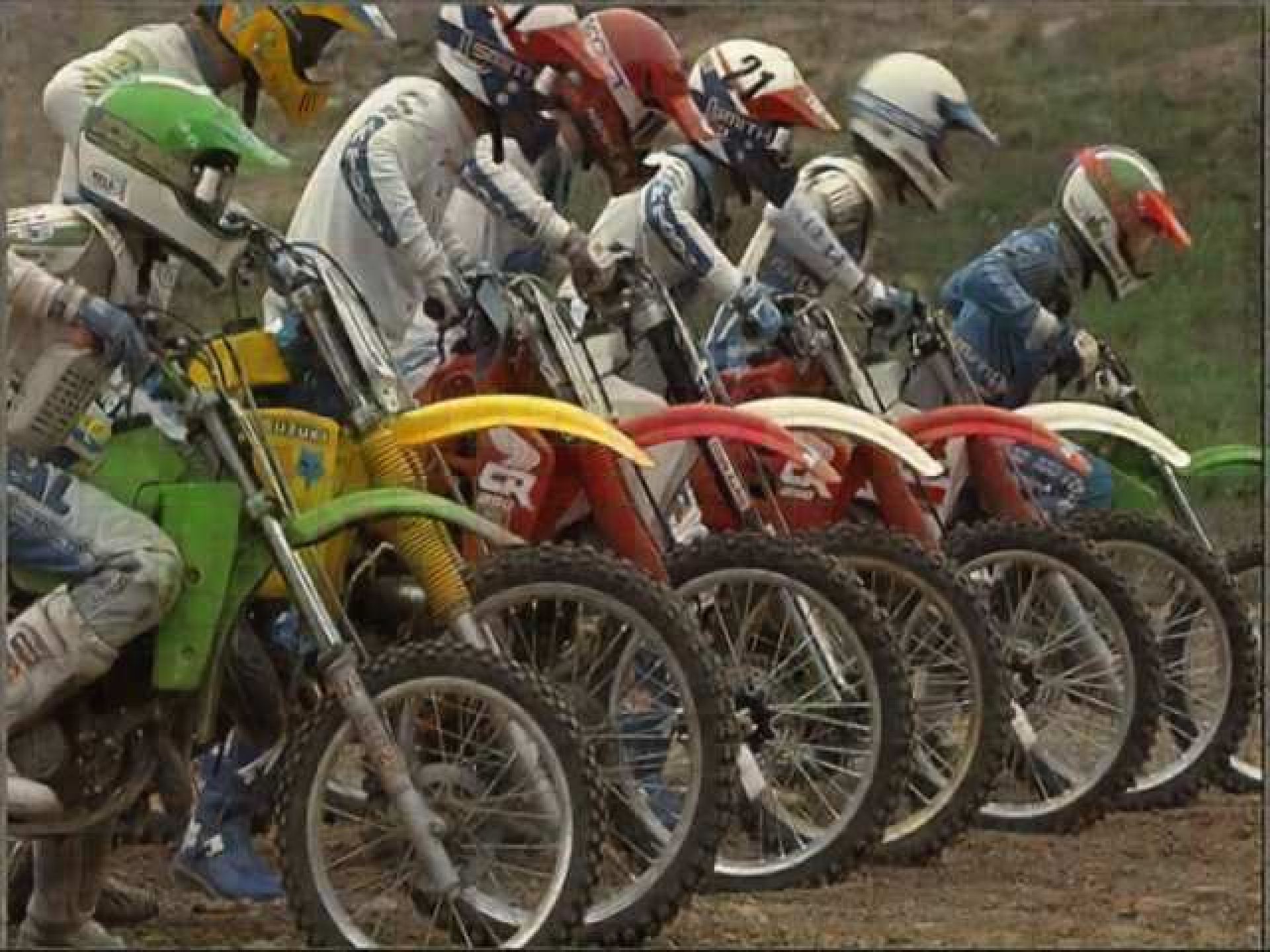}  &
		\includegraphics[width=0.15\textwidth]{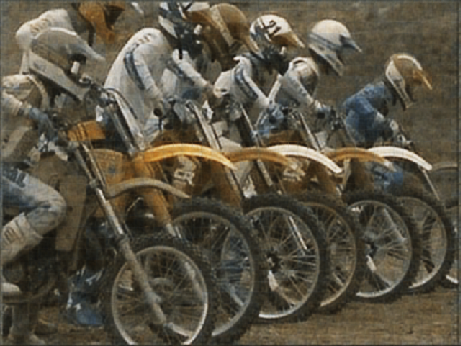} &
		\includegraphics[width=0.15\textwidth]{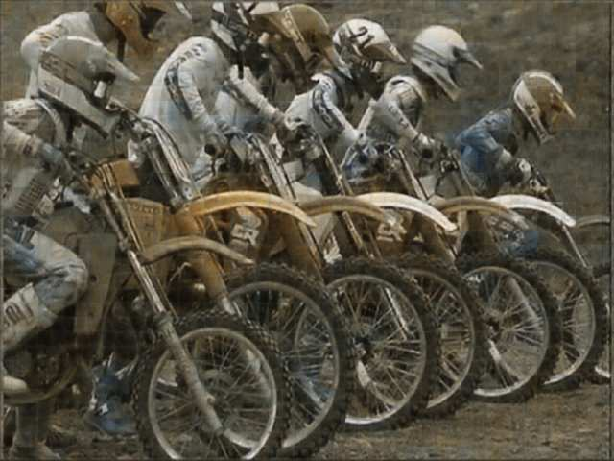} &
		\includegraphics[width=0.15\textwidth]{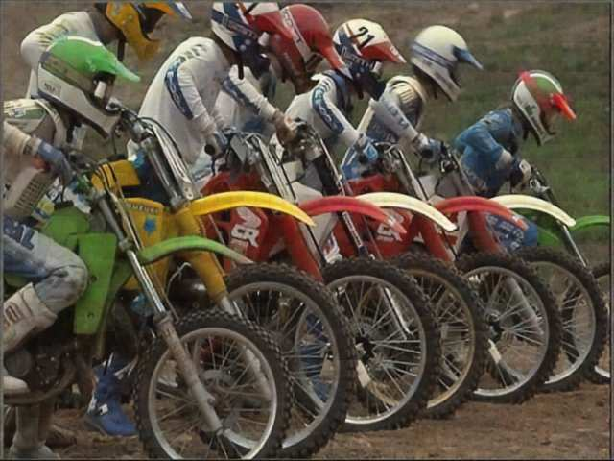}&
		\includegraphics[width=0.15\textwidth]{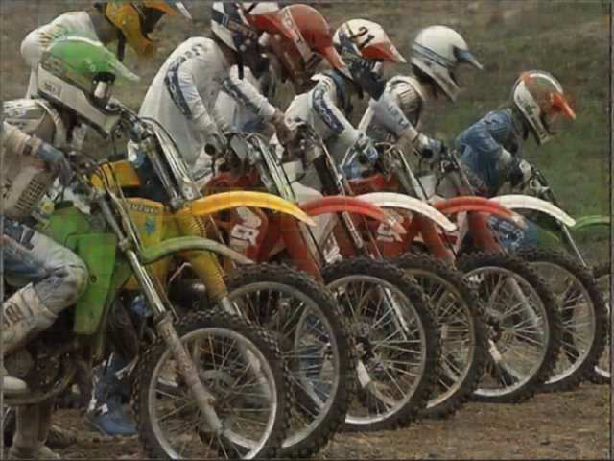}&
		\includegraphics[width=0.15\textwidth]{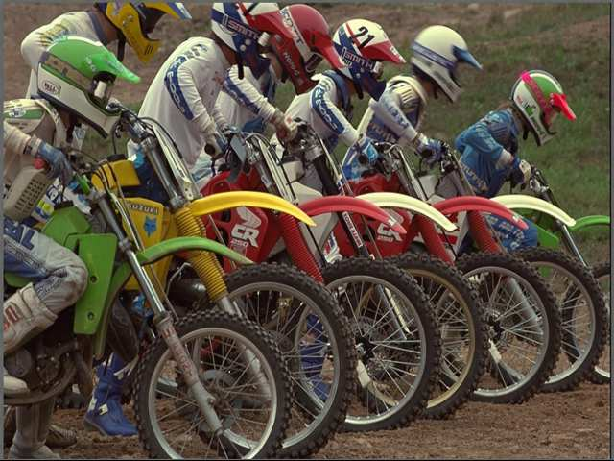}
		\\
		\small{25.97dB/0.85} & 
		\small{21.42dB/0.60} &
		\small{21.73dB/0.69} &
		\small{25.43dB/0.79} &
		\small{23.63dB/0.76} &
		\small{41.59dB/0.99} 
		
		\\
		
		\includegraphics[width=0.15\textwidth]{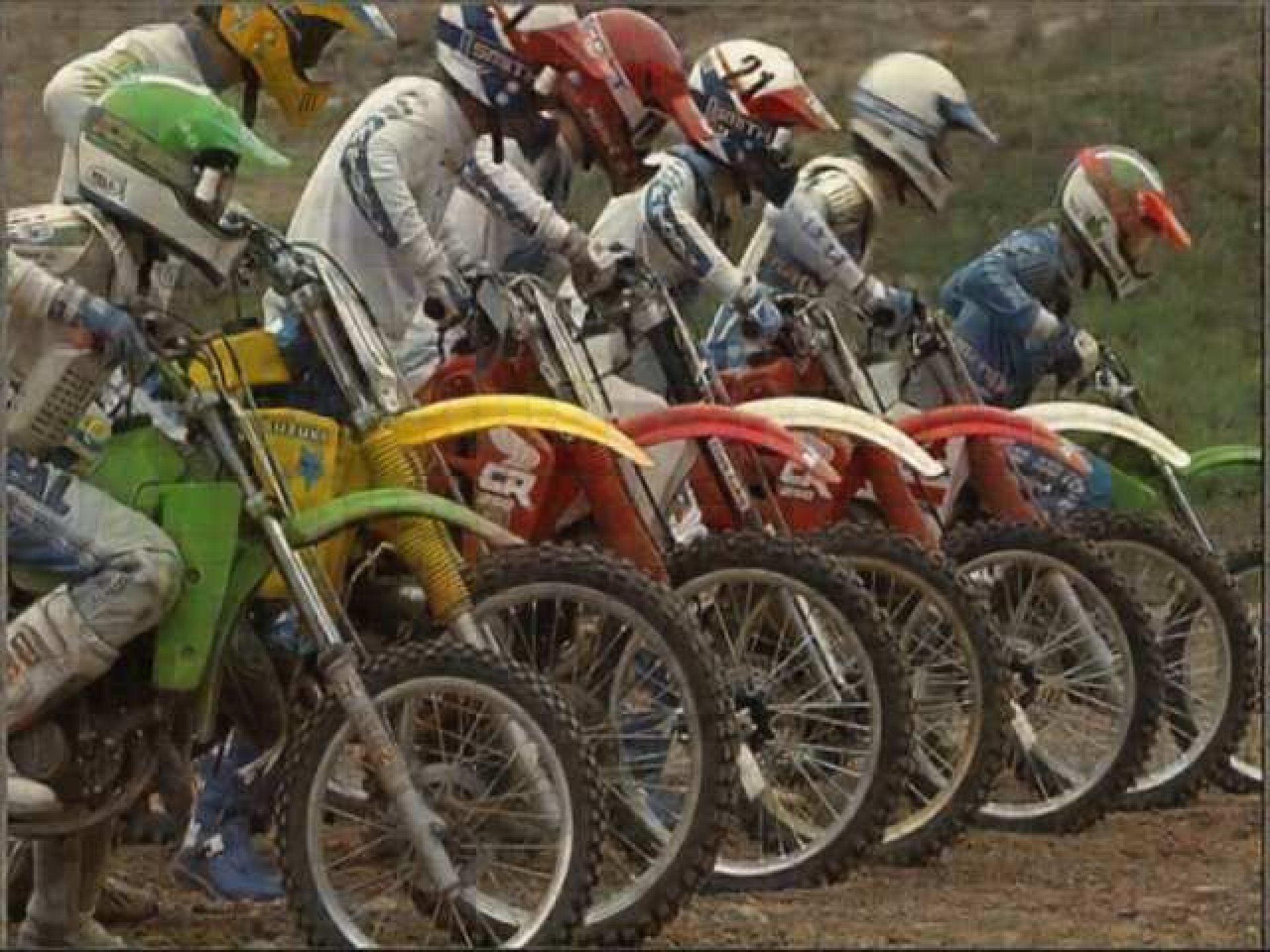}  &
		\includegraphics[width=0.15\textwidth]{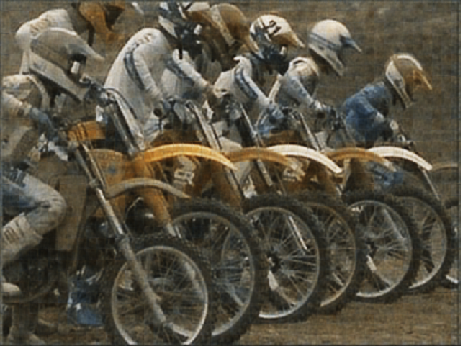} &
		\includegraphics[width=0.15\textwidth]{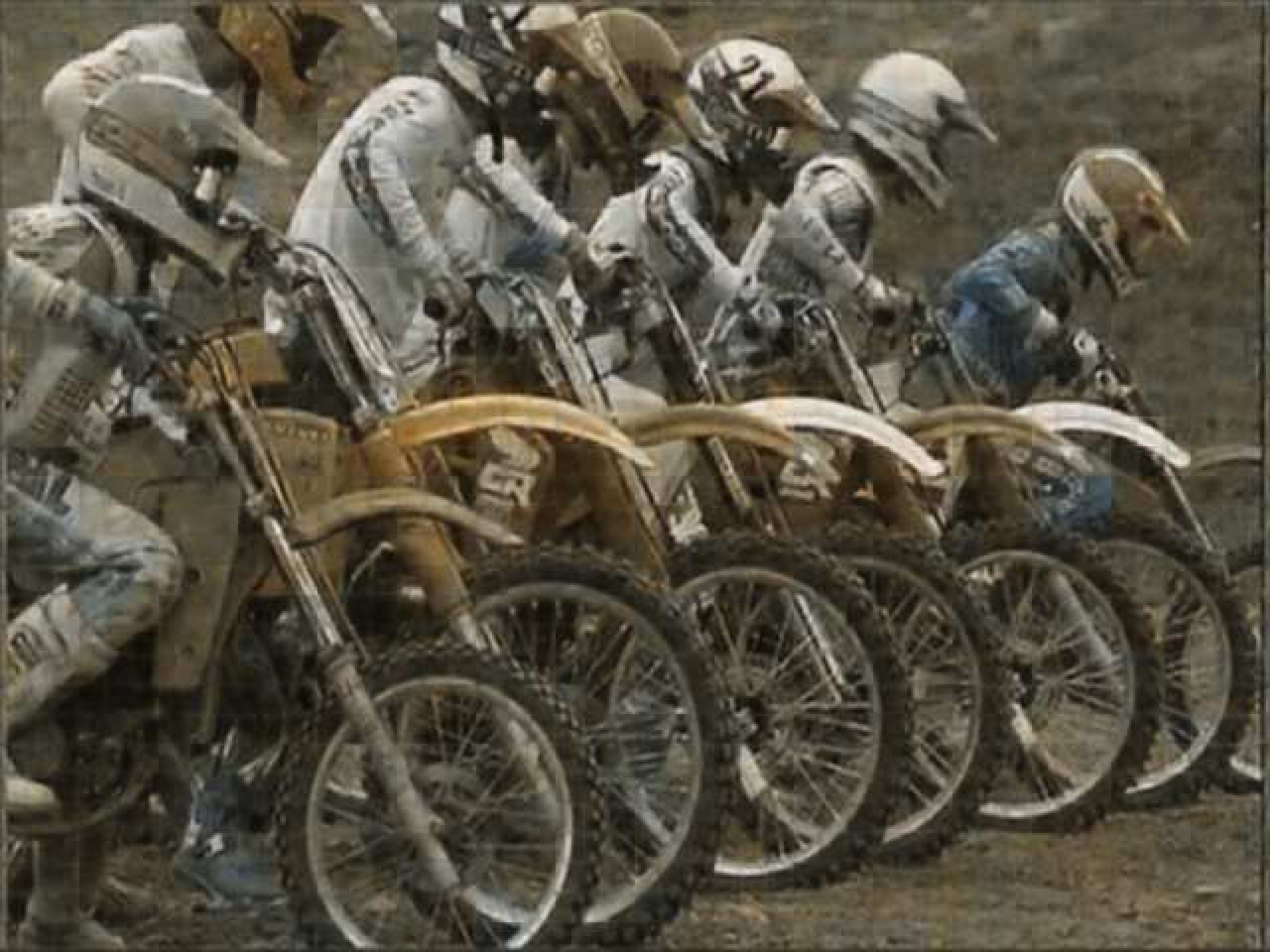} &
		\includegraphics[width=0.15\textwidth]{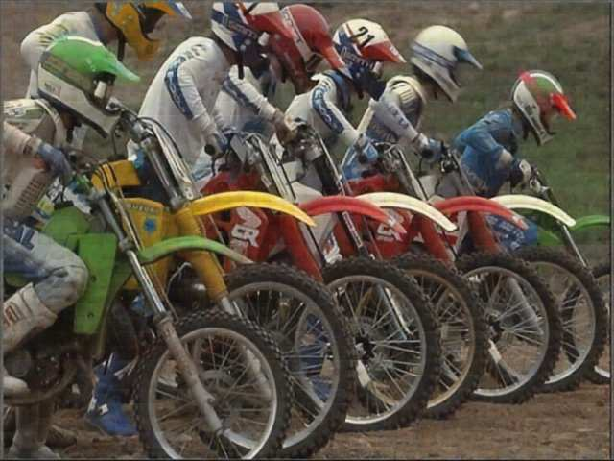}&
		\includegraphics[width=0.15\textwidth]{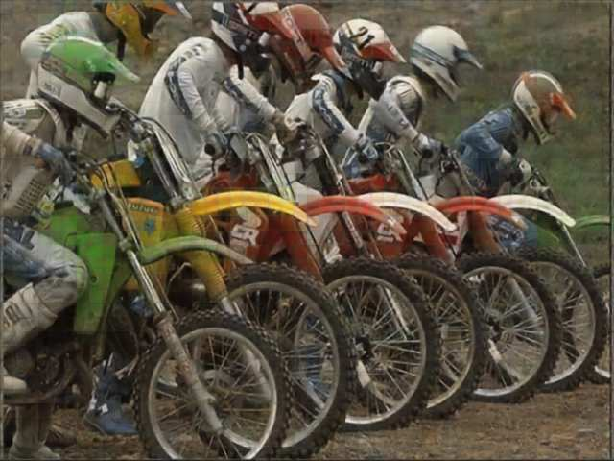}&
		\includegraphics[width=0.15\textwidth]{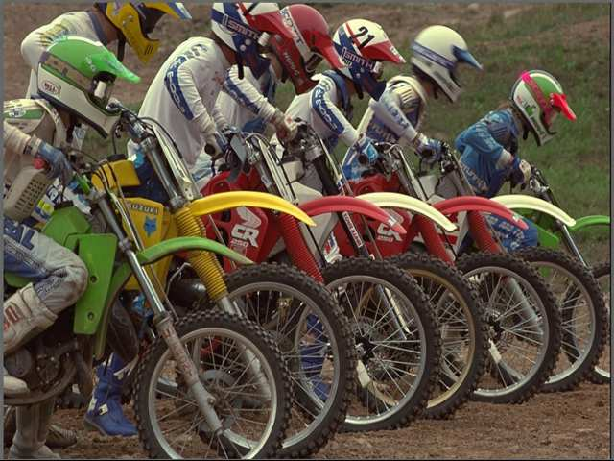}
		\\
		\small{25.39dB/0.83} &
		\small{21.36dB/0.60} & 
		\small{21.75dB/0.77} &
		\small{24.78dB/0.76} &
		\small{23.58dB/0.69} & 
		\small{39.57dB/0.99}
		
		\\

		\includegraphics[width=0.15\textwidth]{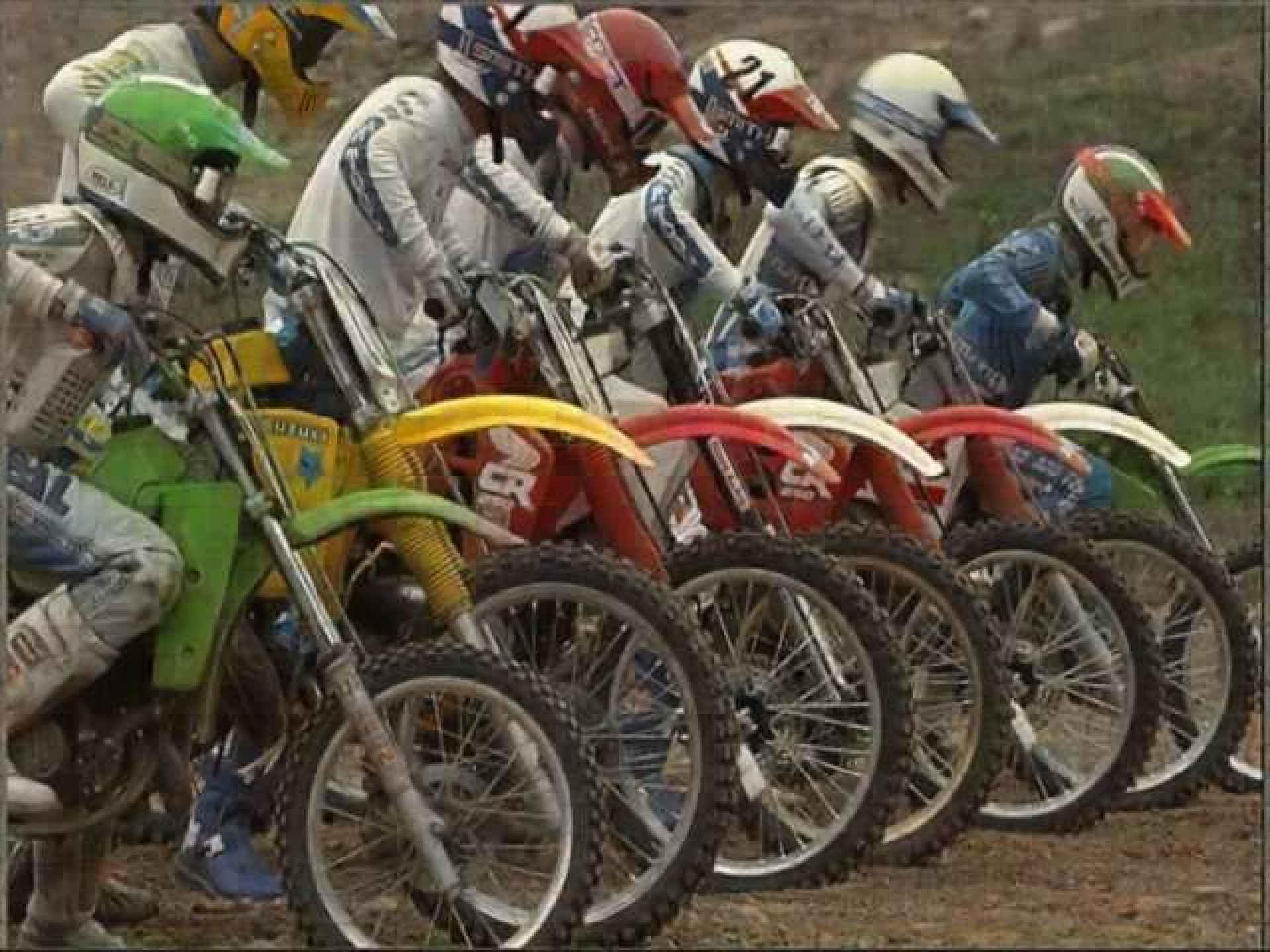}  &
		\includegraphics[width=0.15\textwidth]{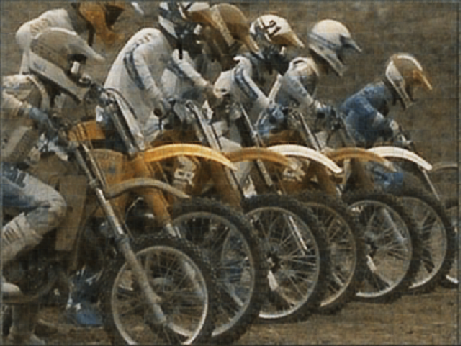} &
		\includegraphics[width=0.15\textwidth]{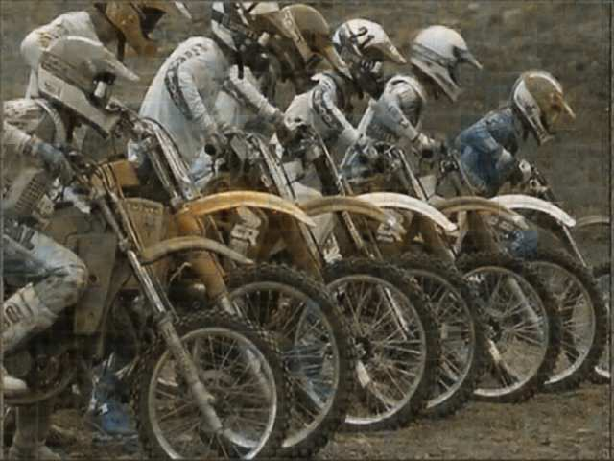} &
		\includegraphics[width=0.15\textwidth]{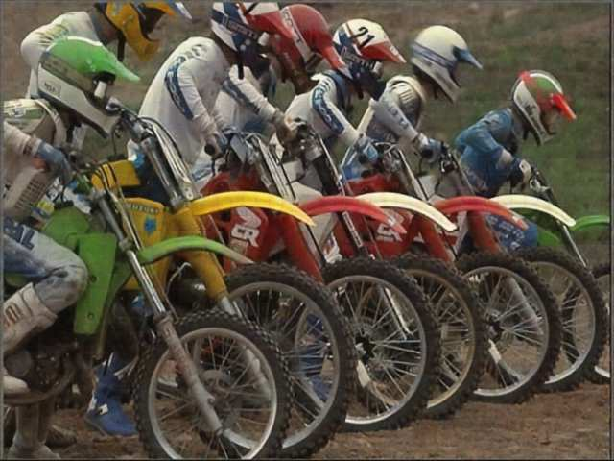}&
		\includegraphics[width=0.15\textwidth]{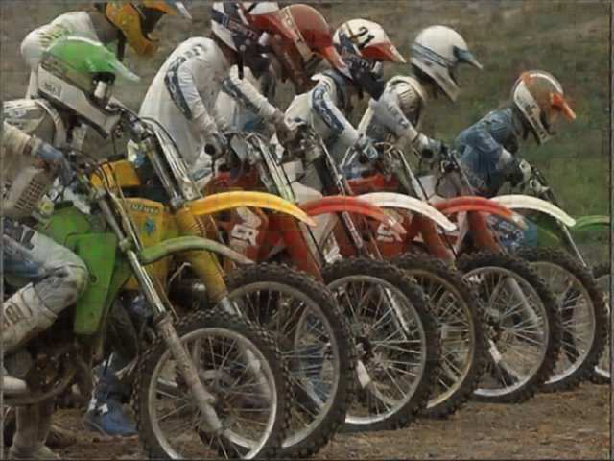}&
		\includegraphics[width=0.15\textwidth]{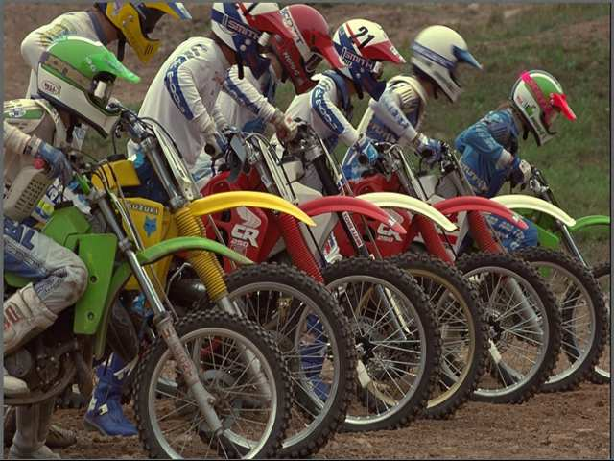}
		\\
		\small{25.95dB/0.85} &
		\small{21.43dB/0.60} & 
		\small{21.74dB/0.69} & 
		\small{25.44dB/0.79} &
		\small{23.62dB/0.76} &
		\small{41.63dB/0.99}
		
	\end{tabular}
	\vspace{0mm}
	\caption{{Generalization performance of our  scheme (pre-trained on CIFAR-10 dataset) evaluated over Kodak dataset and compared with different benchmarks.  
    PSNR/SSIM metrics are reported below each image.}}
	\label{fig:visual_kodak_pic3}
	\vspace{0mm}
\end{figure*}		

\appendices 
\section{\textcolor{black}{Training Process and Ablation Study}}\label{app_TP}

Fig. \ref{fig: TP} demonstrates the training process. Three main performance metrics for training, i.e., the cross-entropy (averaged over all training Eves), SSIM, and MSE are examined. 

The 
	cross-entropy results   
	show that during our \emph{fair minimax training} between the legitimate neural pair    and the adversarial DNNs,  our proposed scheme gradually learns to minimize the sensitive information  leakage, as the average cross-entropy is increasing over episodes. 
    The figure also shows that the training process has been successful in restricting adversaries, with respect to information leakage,  as the training steps go on.   
Our ablation studies in Fig. \ref{fig: TP}  highlight  the importance of loss function (LF) in controlling  the secrecy performance against adversarial DNNs.   This can be seen by investigating the red, green and blue curves, which show almost similar cross-entropy values for the same LF but different DNNs.

Fig. \ref{fig: TP} also demonstrates data reconstruction performance during training, highlighting  the outperformance of our  approach.  This is because  we take into account the perceptual characteristics of image both in the implemented  DNN architecture ($\{\mathsf{Conv2d} + \mathsf{GDN}\}$ units), as well as the proposed LF (considering the SSIM metric).  
	The figure highlights the importance of DNN architecture in the achievable reconstruction performance.   Notably, by implementing the neural architecture  of \cite{AE-Deniz} while considering our proposed LF, the reduction in reconstruction quality in terms of SSIM  is three times more  than the case in which we use our proposed DNN architecture with the LF of \cite{AE-Deniz}.   

\section{Further Experiments on Kodak Dataset}\label{Kodak}
Fig. \ref{fig:visual_kodak_pic3} provides a visual comparison on the generalization performance of our system  over Kodak dateset and with different benchmarks.  For this experiment, channel SNRs are set to  $\Gamma_{E}=15$ dB and $\Gamma_{L}=20$ dB. 
From top to bottom, each row corresponds to the Rayleigh-fading, AWGN, and Nakagami-$m$ channel models, respectively.  
Benchmark 1 corresponds to the scheme proposed in \cite{AE-Deniz}. Benchmark 2 corresponds to the DNN proposed in \cite{AE-Deniz}, while utilizing  our proposed loss function.   Benchmark 3 visualizes the performance of a system in which our proposed neural encoder-decoder are trained via the training framework proposed of  \cite{AE-Deniz}.  
Benchmark 4 corresponds to the scheme proposed in \cite{Deniz-GDN}. 
Benchmark 5  reflects the case in which the security requirement is neglected by setting $w_i = 0$ and training the encoder-decoder pair without any security constraint. 
The last benchmark indicates  that if there is no security requirement,  
our scheme can  reconstruct images with almost lossless reconstruction over different channel models. It also highlights the security-utility trade-off, showing that for achieving security/privacy-aware deep-JSCC, around $15\%$ decrease in reconstruction performance (in terms of SSIM) can be imposed to the model.   
Fig. \ref{fig:visual_kodak_pic3} validates the outperformance of our  scheme compared with other benchmarks, both visually and numerically.    This can not only be seen from the PSNR/SSIM values, but also from the visual details (e.g., in the first image) and the color combinations (e.g., in the second image) of the reconstructed images. 
Comparing the PSNR/SSIM values of benchmarks 1 to 4 highlights the importance of implementing well-designed DNN architecture for the neural encoder-decoder pairs, which seems to have more impact on the reconstruction performance compared to the  employed loss function.


\end{document}